\documentclass[a4paper,11pt]{article}
\pdfoutput=1 

\usepackage{jcappub} 

\usepackage[T1]{fontenc} 
\usepackage{fullpage}
\usepackage{CJKutf8} 
\usepackage[utf8]{inputenc}
\usepackage{setspace}
\usepackage{parskip}
\usepackage{titlesec}
\usepackage[section]{placeins}
\usepackage{xcolor}
\usepackage{breakcites}
\usepackage{lineno}
\usepackage{textcomp}
\usepackage{epstopdf}
\usepackage{hyphenat}

\usepackage{enumitem}
\usepackage{natbib}
\usepackage{multicol}

\title{\boldmath Influence of Fermionic Dark Matter on the Structural and Tidal Properties of Neutron Stars}

\author[]{Monmoy Molla,}
\author[]{Masum Murshid,}
\author[1]{Mehedi Kalam\note{Corresponding author}.}

\affiliation[]{Department of Physics, Aliah University, New Town, Kolkata, 700160, West Bengal, India}

\emailAdd{monmoymolla2016@gmail.com}
\emailAdd{masum.murshid@wbcte.ac.in}
\emailAdd{mehedikalam.phys@aliah.ac.in}

\abstract{We investigate the influence of ideal Fermi gas dark matter on the observable properties of neutron stars (NSs). Our analysis considers dark matter (DM) particle masses ($\mu$) ranging from $0.2$ GeV to $1$ GeV and various DM mass fractions ($f$). By examining the coexistence of DM and baryonic matter (BM), we explore the formation of either a dense DM core or an extended dark halo within NSs. Our findings indicate that the resulting DM distribution depends critically on both $\mu$ and $f$. We systematically explore the parameter space of the fermionic DM model using two representative BM equations of state (EoSs) by applying constraints from NS radius measurements by the Neutron Star Interior Composition Explorer (NICER), observations of $2M_{\odot}$ NSs, and tidal deformability limits from the LIGO/Virgo Collaboration. This comprehensive analysis enables us to exclude specific ranges of $\mu$ and $f$, demonstrating that the amount of accumulated DM must be relatively small to satisfy current astrophysical constraints.

}

\begin{document}
\maketitle
\flushbottom

\section{Introduction}
\label{874460}
Dark matter (DM) is widely recognized as comprising the majority of the universe's mass, yet its fundamental characteristics remain elusive. Multiple observational phenomena provide compelling evidence for DM's existence, including the flat rotation curves observed in spiral galaxies \citep{sofue2001rotation}, gravitational lensing observations \citep{massey2010dark}, and anisotrpies in the cosmic microwave background radiation \citep{hu2002cosmic}. Among the leading theoretical candidates for DM particles are Weakly Interacting Massive Particles (WIMPs), axions, and sterile neutrinos \citep{bertone2018new,bertone2018history}. Numerous major experimental collaborations, such as XENONIT \citep{aprile2021search}, LUX-ZEPLIN \citep{raaijmakers2019nicer}, and the China Dark Matter Experiment (CDEX) \citep{dai2022exotic}, are currently conducting searches to directly detect DM particles. Determining the precise identity and properties of DM would represent a major breakthrough in fundamental physics.

Given that all current observational evidence for DM comes from gravitational effects, it is plausible that gravity represents the sole means of interaction between DM and baryonic matter (BM). However, the inherently weak nature of gravitational coupling makes it challenging to probe DM through its interactions with BM. One promising approach is to examine DM on cosmological scales, where it contributes substantially to the overall gravitational field. Alternatively, investigating high-density environments where gravity plays a dominant role offers another viable strategy. In this context, compact objects such as  neutron stars (NSs) serve as ideal natural laboratories for studying DM properties.

The extreme matter densities present in NS cores mean that the underlying physics of these regions remains poorly understood. Consequently, investigating NS properties is both crucial and compelling, as such studies can help constrain the uncertain equations of state (EoSs) of nuclear matter. While deriving nuclear matter EoS from first principles remains beyond current capabilities, significant constraints have emerged from both nuclear physics experiments and NS observations. For instance, precise measurements of the neutron skin thickness of $^{208}$Pb have constrained the density dependence of symmetry energy near saturation density \citep{adhikari2021accurate}. Mass determinations of NSs through Shapiro delay measurements or orbital period analysis of binary systems have revealed some of the most massive NSs known, including PSR J0348+0432 ($2.01\pm 0.04M_{\odot}$) and PSR J0740+6620 ($2.14^{+0.10}_{-0.09}M_{\odot}$) \citep{antoniadis2013massive,cromartie2020relativistic}. More recently, Keck telescope optical spectrophotometry and imaging of companions to PSR J1810+1744 and PSR J0952-0607 have yielded remarkable mass measurements of $2.13\pm0.04M_{\odot}$ and $2.35\pm0.17M_{\odot}$, respectively \citep{romani2021psr,romani2022psr}. These observations have already excluded numerous soft EoS models. Additionally, gravitational wave (GW) data from the LIGO/Virgo collaboration constrained the tidal deformability of a $1.4M_{\odot}$ NS ($\Lambda_{1.4}$) to be below 580 based on the GW170817 event \citep{abbott2018gw170817}. The detection of a $2.6M_{\odot}$ compact object in GW190814 \citep{abbott2020gw190814} poses an additional challenge to our understanding of dense matter if confirmed as a NS. Moreover, the Neutron Star Interior Composition Explorer (NICER) X-ray telescope has provided joint precision measurements of mass and radius for PSR J0030+0451, PSR J0740+6620, PSR J0437-4715 and, PSR J0614-3329, opening new avenues for understanding the structure of NSs \citep{miller2019psr,riley2019nicer,miller2021radius,riley2021nicer,choudhury2024nicer,mauviard2025nicer}. With anticipated increases in NS observations through both electromagnetic (EM) and GW channels, significant progress in constraining the nuclear matter EoS is expected in the coming decade. Beyond nuclear physics, NSs also offer a unique opportunity to probe DM properties and address one of the fundamental questions in contemporary physics.

Astrophysical objects comprising mixtures of fermions and bosons that interact solely through gravity have been investigated using two distinct approaches: (1) Fermion-Boson stars \citep{valdez2013dynamical,valdez2020fermion,nyhan2022dynamical,jockel2024fermion} and (2) DM-admixed NSs \citep{ellis2018dark,nelson2019dark,rafiei2022bosonic,leung2022tidal,thakur2024feasibility} (DANSs). Computing equilibrium configurations of fermion-boson stars requires solving the coupled Einstein-Klein-Gordon equations \citep{liebling2023dynamical}. \textcolor{blue}{In addition, state-of-the-art Bayesian inference methods have recently been employed on actual GW inspiral observations, such as GW230529, to place observational constraints on bosonic DANSs. \citep{santos2025observational}}. For DANSs, both fermions and bosons are treated as perfect fluids, with the two-fluid Tolman-Oppenheimer-Volkoff (TOV) formalism employed to describe these mixed compact objects. Three distinct scenarios can emerge for DANSs where BM and DM interact exclusively via gravity \citep{li2012gravitational,mukhopadhyay2017compact,rafiei2022bosonic}:

\begin{enumerate}[label=(\roman*), leftmargin=2.5em]
    \item \textbf{Core-condensed DM}: DM accumulates in a compact core at the NS center, with the DM radius $R_D$ smaller than the baryonic radius $R_B$, such that $R_B > R_D$.
    
    \item \textbf{Uniformly distributed DM}: DM is distributed throughout the entire NS, yielding identical radii for both components: $R_B = R_D$.
    
    \item \textbf{Extended DM halo}: DM forms an extended envelope around the NS, with the DM radius exceeding the baryonic radius: $R_D > R_B$.
\end{enumerate}

Numerous studies demonstrate that DM presence can significantly modify the NS structure and thermodynamic properties, producing detectable signatures in astrophysical observations \citep{ellis2018dark,nelson2019dark,rafiei2022bosonic,leung2022tidal,thakur2024feasibility,panotopoulos2017dark,das2022dark,liu2024dark,shakeri2024bosonic}. These investigations reveal that DM incorporation generally softens the NS EoS, with profound implications for the mass(M)-radius(R) relation. For heavier DM particles, the maximum gravitational mass($M_{max}$) decreases substantially, potentially rendering the $2M_{\odot}$ threshold unattainable, whereas lighter DM particles can induce a transition from a dark core to a halo configuration, occasionally enabling higher $M_{max}$ \citep{rafiei2022bosonic,leung2022tidal,shakeri2024bosonic}. Such effects are critical for interpreting observational constraints. These studies also consistently demonstrate that DM reduces the dimensionless tidal deformability ($\Lambda$) of NSs \citep{rafiei2022bosonic,leung2022tidal,liu2024dark}. Here a particularly noteworthy result is that the $\Lambda$ exhibits an abrupt transition from pure BM star behavior to pure DM star behavior within a narrow range of intermediate DM mass fractions. This characteristic feature can be exploited to constrain the DM model parameter space and the DM content within NSs in accordance with observational constraints \citep{ellis2018dark,nelson2019dark,zhang2020constraint,sen2021implications,das2019confronting}. Furthermore, multimessenger observations combining GW detections (from facilities such as LIGO/Virgo/KAGRA \citep{abbott2021observation}) with X-ray measurements (from instruments like NICER \citep{miller2019psr,raaijmakers2020constraining}) provide powerful tools for probing the presence of DM within or surrounding NSs \citep{silva2021astrophysical}.

In this study, we investigate the impact of ideal fermionic DM on NS properties, including $M_{max}$, $R$, and  $\Lambda$ across different DM distribution configurations within a two-fluid framework. We adopt the assumption that DM and BM interact exclusively through gravitational coupling. Our analysis focuses on DM particle masses ($\mu$) ranging from $0.2$ GeV to $1$ GeV. The DM component is modeled as a zero-temperature ideal Fermi gas governed by the EoS formulated by Oppenheimer and Volkoff \citep{oppenheimer1939massive}. To account for uncertainties in nuclear EoSs, we employ the stiff MPA1 and soft FSU2R EoSs, which represent different limiting values for $M_{max}$, $R$, and $\Lambda$.

We demonstrate that the formation of either a DM halo or DM core depends on the DM particle mass ($\mu$) and  DM mass fraction ($f$). Subsequently, we examine how DM core/halo formation affects the $M_{max}$, visible radius ($R_B$), dark radius ($R_D$), and $\Lambda$ in the context of current observational constraints, namely the maximum mass requirement $M_{\rm max}\gtrsim 2M_{\odot}$, the radius constraint $R_{1.4}\gtrsim 11$ km, and the tidal deformability limit $\Lambda_{1.4}\lesssim 580$. Our results indicate that fermionic DM forming a NS core leads to reductions in both R and $M_{max}$ of the DANS, which is inconsistent with recent measurements. Additionally, DM halo formation substantially enhances $\Lambda$, conflicting with the latest observational bounds. By combining EM and GW constraints, we determine the permitted DM model parameter space through systematic scanning of $\mu$ and $f$. This analysis establishes upper limits on $f$ within NSs for our considered mass range ($0.2-1$ GeV) using both stiff and soft BM EoSs.

The manuscript is structured as follows. Section ~\ref{507127} presents the EoSs for both BM and fermionic DM. Section ~\ref{222222} outlines the formalism for computing hydrostatic equilibrium configurations and tidal parameters of DANSs. Section ~\ref{44444} analyzes the influence of fermionic DM on the mass ($M$), $R$, and $\Lambda$ of NSs. Section ~\ref{77777} provides a systematic parameter space exploration of $\mu$ and $f$ in relation to three critical observational constraints: $M_{\rm max}\gtrsim 2M_{\odot}$, $R_{1.4}\gtrsim 11$ km, and $\Lambda_{1.4}\lesssim 580$. Section ~\ref{88888} summarizes our findings. Throughout this work, we adopt natural units where $G=\hbar = c = 1$.

\section{BM and DM models}
\label{507127}
\subsection{\textbf{Equation of State for BM}}
To model BM, we employ two well-established hadronic EoSs. The first is based on a relativistic mean-field description with $\sigma$--$\omega$--$\rho$ meson exchange, using the ``FSU2R'' parametrization \citep{tolos2017equation}. In this model, nucleon-nucleon interactions are mediated by the exchange of $\sigma$ and $\omega$ mesons, while the inclusion of $\rho$-meson couplings accounts for isovector properties and reproduces nuclear saturation. This EoS yields a NS with $M_{max}$ of $2.05,M_{\odot}$, $R_{1.4} = 12.8~\mathrm{km}$, and  $\Lambda_{1.4} = 622$. It satisfies modern astrophysical constraints---including the $2\,M_{\odot}$ mass threshold, recent radius measurements below $13~\mathrm{km}$, $\Lambda$ measurement below $580$ as discussed in  previous section.The EoS for the outer crust part has been taken from BPS EoS \citep{baym1971ground} whereas for the inner crust part self consistent Thomas-Fermi approximation has been applied \citep{sharma2015unified}.

In addition, we consider a stiffer nuclear EoS derived from the relativistic Brueckner--Hartree--Fock (RBHF) approach \citep{muther1987nuclear}, denoted as MPA1. Here, the nucleonic interaction is described through the exchange of $\pi$ and $\rho$ mesons. This stiff EoS produces a neutron star with $M_{max}$ of $2.4\,M_{\odot}$, $R_{1.4} = 12.6~\mathrm{km}$, and $\Lambda_{1.4} = 516$, which likewise conforms to the current observational bounds discussed in Section~~\ref{874460}.
\subsection{\textbf{Equation of State for DM }}
\label{111111}
The exact nature of DM has not been properly understood until now and is a hot topic of research. We have used the zero-temperature ideal Fermi gas EoS for the DM EoS, considering the DM particles to be fermionic \citep{oppenheimer1939massive}.The DM EoS considered here is
\begin{equation}
    \rho = J(sinhx - x),
\end{equation}
\begin{equation}\,
    p = \frac{1}{3}\, J\, (3x + sinhx - 8 sinh\frac{1}{2}x),
\end{equation}
where 
\begin{equation}
    J = \frac{\pi \mu^4}{4h^3},
\end{equation}
\begin{equation}
    x = 4\,ln\,[(1+y^2)^\frac{1}{2} + y],
\end{equation}
and 
\begin{equation}
y = \frac{1}{\mu}\left(\frac{3nh^3}{8\pi}\right)^\frac{1}{3}    
\end{equation}
Here $\textit{n}$ is the number density of DM particles. $\mu$ denotes DM particle mass and $y$ is related to the momentum of the particle distribution. 

\section{Theoretical Formalism}
\label{222222}
\subsection{\textbf{Hydrostatic Configuration}}
\label{sec:2}
\indent The  form of metric that describes a spherically symmetric, static spacetime is
\begin{equation}
ds^2= -e^{\nu(r)}dt^2 + e^{\lambda(r)}dr^2 + r^2 d\theta^2 + r^2\sin^2{\theta}d\phi^2    
\end{equation}
The equation for a static, spherically symmetric, non-rotating compact star can be obtained from this metric, and this equation is called the Tolman-Oppenheimer-Volkoff equation(TOV)\citep{oppenheimer1939massive}:
\begin{equation}
\frac{dp}{dr}=- \frac{M(r)+4\pi r^3 p(r)}{r^2 (1-\frac{2M(r)}{r})}(\rho(r) + p(r))
\label{eq:corrected1}
\end{equation}
Also the equation that describes how the mass enclosed 
at a given radius r changes as we move outward from the center of the star is 
\begin{equation}
 \frac{dM}{dr}=4\pi r^2 \rho(r) 
 \label{eq:corrected2}
\end{equation}
Here $M(r)$ represents the mass of the compact object surrounded by a radial distance, $r$ and its energy density at distance $r$ is $\rho(r)$.The metric function $\lambda(r)$ also relates $M(r)$ through the relation $e^{-\lambda(r)}=1-\frac{2M(r)}{r}$.The derivative of the metric function $\nu(r)$ can be connected with the star's mass, radius and pressure by the relation
\begin{equation}
\frac{d\nu}{dr}=\frac{2(M(r)+4\pi p(r)r^3)}{r^2 (1-\frac{2M(r)}{r})}
\end{equation}
The boundary conditions for the equations (\ref{eq:corrected1}) and (\ref{eq:corrected2}) are at the center of the star, its mass is zero, i.e.,$M(r=0)=0$ and star has a density at its center called central density, $\rho(r=0)=\rho_{c}$. It can be shown that $e^{\nu(R)}=1-\frac{2M_{T}}{R}$ is satisfied on the surface of the star by the metric function $\nu(r)$, where $M_{T}\,=\,M(r=R)$ is the star's total mass and $R$ is its total radius.
\indent For the study of DANS using the two-fluid model, we have to modify the TOV equation a little bit. One of our presumptions is that BM and DM only interact gravitationally with one another.Despite simplifying the intricate system, this gravitational-only coupling poses a significant theoretical constraint. Neglecting possible nongravitational interactions, such as kinetic friction, momentum transfer, or direct interactions via mediator particles, in the extremely high density core of an NS could result in an overestimate of the pressure support that the DM component provides. \textcolor{blue}{In this context, recent work has explored DM-nucleon interactions mediated by the Higgs boson, showing that such non-gravitational couplings can significantly influence nuclear saturation properties and the stellar EoS \citep{kumar2024constraints}. } Therefore, the total energy density can be expressed in the form 
\begin{equation}
\rho(N_{BM},N_{DM})=\rho_{BM}(N_{BM})+\rho_{DM}(N_{DM})
\end{equation}
Like energy density, the expression of pressure also includes two parts: the pressure of BM and that of DM \citep{ciarcelluti2011have}.
\begin{equation}
p(N_{BM},N_{DM})=p_{BM}(N_{BM})+p_{DM}(N_{DM})
\end{equation}
We will solve the TOV equation here using the two-fluid model. In this model it is assumed that the DM and BM are interacting with each other only via gravity. The TOV equation and mass equation in this model are \citep{ciarcelluti2011have,leung2022tidal,Vikiaris:2023jol}:
\begin{equation}
\label{eq:first}
 \frac{dp_{BM}}{dr}=-(p_{BM}+\rho_{BM})\frac{d\nu}{dr}   
\end{equation}
\begin{equation}
\label{eq:sec}
\frac{dM_{BM}}{dr}=4\pi r^2 \rho_{BM}
\end{equation}
\begin{equation}
\label{eq:third}
 \frac{dp_{DM}}{dr}=-(p_{DM}+\rho_{DM})\frac{d\nu}{dr}   
\end{equation}
\begin{equation}
\label{eq:four}
\frac{dM_{DM}}{dr}=4\pi r^2 \rho_{DM}
\end{equation}
\begin{equation}
\frac{d\nu}{dr}=\frac{(M_{BM}+M_{DM})+4\pi r^3(p_{BM}+p_{DM})}{r(r-2(M_{BM}+M_{DM}))}
\end{equation}
 The systems of above equations will be solved along with differential equations for the  tidal deformability and  second love number $k_{2}$ . The boundary conditions of the above system of equations are 
\begin{equation}
M_{BM}(r=0)=0
\end{equation}
\begin{equation}
\rho_{BM}(r=0)=\rho_{c,BM}
\end{equation}
\begin{equation}
M_{DM}(r=0)=0
\end{equation}
\begin{equation}
\rho_{DM}(r=0)=\rho_{c,DM}
\end{equation}
Given the fixed central pressure values $p_{c,\mathrm{DM}}$ and $p_{c,\mathrm{BM}}$, and with the initial masses $M_{\mathrm{DM}}(r=0) = M_{\mathrm{BM}}(r=0) = 0$, we solved the system of equations (Eqs.~\eqref{eq:first}--\eqref{eq:four}) via numerical integration. \textcolor{blue}{The integration was terminated at the radial coordinate where the pressure of one of the components becomes zero. This specifies either the DM core radius, $R_{D}$, or the BM core radius, $R_{B}$.}
\textcolor{blue}{Now in the two-fluid framework, the boundary conditions at the star's surface are defined separately for each component:
\begin{equation}
 p_{\rm BM}(r=R_B) = 0  
\end{equation}
\begin{equation}
 p_{\rm DM}(r=R_D) = 0   
\end{equation}
The BM pressure, $p_{BM}$, drops to zero at the visible surface of the NS, that is, at $r = R_B$, while the DM pressure, $p_{DM}$, becomes zero at the DM radius $R_D$.}
Consequently, in scenarios (i) and (ii) (see Introduction section), for which DM resides inside the NS ($R_{B} \geq R_{D}$), the numerical integration is extended to the radius where the BM pressure drops to zero, \textcolor{blue}{$p_{BM}(r=R_{B}) = 0$}, defining the visible stellar surface. The total gravitational mass is then defined as
\begin{equation}
\label{eq:five}
    M_{T} = \int_{0}^{R_{B}} 4\pi r^{2} \left[ \rho_{BM}(r) + \rho_{DM}(r) \right] dr.
\end{equation}
In the case of a NS embedded within an extended DM halo (scenario (iii) in Introduction section), characterized by $R_D > R_B$ and vanishing baryonic pressure beyond \textcolor{blue}{$r\geq R_B$}, the total gravitational mass is calculated by replacing the upper limit of integration in Eq.(~\ref{eq:five}) with the DM radius $R_D$. Consequently,
\begin{equation}
    M_T = M_{BM}(R_B) + M_{DM}(R_D),
\end{equation}
where $M_{BM}(R_B)$ and $M_{DM}(R_D)$ are the integrated BM and DM mass contributions, respectively. The observable radius of the star, however, is still $R_B$, as the baryonic surface is electromagnetically visible, whereas the DM halo radius $R_D$ is not directly detectable with current methods. \textcolor{blue}{An important point is that, in the DM core configuration ($R_B \geq R_D$), the total pressure, defined as $p(r)=p_{BM}(r)+p_{DM}(r)$, drops to zero at the outermost surface of the star (i.e., $p(r=R_B=R)=0$ in this configuration), where $R$ represents the total, or outer, stellar radius. In contrast, for the DM halo configuration ($R_D > R_B$), the total pressure likewise vanishes at the outer boundary (i.e., $p(r=R_D=R)=0$ in the DM halo case), but it remains finite at the visible surface $r=R_B$ (that is, $p(r=R_B)\neq 0$ in the DM halo configuration).}
Another key parameter in our analysis is the DM mass fraction, $f$, which quantifies the proportion of DM within a DANS. It is defined as
\begin{equation}
    f = \frac{M_{DM}(R_{D})}{M_{T}},
\end{equation}
where $M_{DM}(R_{D})$ is the total DM mass enclosed within its radius $R_D$, and $M_T$ is the total gravitational mass of the DANS.
\subsection{\textbf{The Tidal Deformability and Tidal Love Number}}
The newest addition to astrophysical methods for probing properties of NSs focuses on examining their response to the tides. The external perturbing tidal gravitational field produced by its own companion star distorts the shape of the NS. This change in shape is measured by a parameter known as tidal deformability $\lambda$.Mathematically it can be written as the ratio of the induced quadrupole moment tensor $(Q_{ij})$ to the  external perturbing field tensor $(\epsilon_{ij})$ that generates the perturbation.
\begin{equation}
\lambda=-\frac{Q_{ij}}{\epsilon_{ij}}
\end{equation}
For a non-rotating star, the determination of tidal deformability is given in many articles\citep{hinderer2008tidal,damour2009relativistic,postnikov2010tidal}. The expression of the tidal love number can be written as follows:
\begin{align}
k_2 &= \frac{8}{5} \beta^5 (1-2\beta)^2 [2 - y_R + 2\beta (y_R - 1)] \times \left[ \right. \notag \\
&\quad 2\beta \left( 6 - 3y_R + 3\beta (5y_R - 8) \right) \notag \\
&\quad + 4\beta^3 \left( 13 - 11y_R + \beta (3y_R - 2) + 2\beta^2 (1 + y_R) \right) \notag \\
&\quad + 3 (1-2\beta)^2 \left( 2 - y_R + 2\beta (y_R - 1) \right) \log(1 - 2\beta) \left. \right]^{-1},
\label{eq:love}
\end{align}

Here $\beta=\frac{M_{T}}{R}$ is a parameter called compactness  and $y_{R}=\frac{rH^{\prime}(r)}{H(r)}\left|_{r = R}\right.$, The function $H(r)$ is essentially a solution to the differential equation that follows:
\begin{equation}
H^{\prime\prime}(r)+H^{\prime}(r)F(r)+H(r)Q(r)=0
\end{equation}
where,
\begin{equation}
F(r)=\frac{2}{r}+e^{\lambda(r)}\left(\frac{2M(r)}{r^2}+4\pi r(p(r)-\rho(r))\right)
\end{equation}
\begin{equation}
\begin{aligned}
Q(r)=-\frac{6e^{\lambda}}{r^2}+4\pi e^{\lambda}\left(5\rho(r)+9p(r)+\frac{\rho(r)+p(r)}{(\frac{dp}{d\rho})}\right)-\nu^{\prime^{2}} 
\label{eq:Q}
\end{aligned}
\end{equation}
The primes indicate spatial derivatives $\lambda(r)$ and $\nu(r)$ are the metric functions defined as 
\begin{equation}
e^{\lambda(r)}=\left(1-\frac{2M(r)}{r}\right)^{-1}
\end{equation}
In terms of $y(r)$ the equation $(23)$ taking form:

\begin{equation}
\begin{aligned}
    r y'(r) + y^2(r) &+ y(r) e^{\lambda(r)} \bigl[ 1 + 4 \pi r^2 (p(r) - \rho(r)) \bigr] \\
    &+ r^2 Q(r) = 0,
\end{aligned}
\label{eq:diffeqn}
\end{equation}
The boundary condition of the above equation is 
$y(r=0)=2$. After solving the equation both outside and inside the star, two solutions were found. When the two solutions are matched on the star's surface, the expression for the quadrupolar tidal Love number Eqs(~\ref{eq:love}) was obtained. The inverse length squared is the unit of the perturbing tidal field $\epsilon_{ij}$ \citep{poisson2014gravity}.The quadrupole moment is measured in length cubes. So tidal deformability is written as:
\begin{equation}
    \lambda=\frac{2}{3}k_{2}R^5
\end{equation}
where $k_2$ is the well-known quadrupolar Love number and $R$ is the star's outermost radius.

The dimensionless tidal deformability is also used in many cases. It is defined as 
\begin{equation}
\Lambda \equiv \frac{\lambda}{M_{T}^5}=\frac{2}{3}k_2\frac{R^5}{M_{T}^5}=\frac{2}{3}k_2\beta^{-5}
\label{eq:ok}
\end{equation}
where $\beta$ is the star's compactness and $M_T$ is its total gravitational mass. 

\hspace{1em} For a NS modeled as a two-fluid system consisting of DM and BM, the macroscopic thermodynamic and gravitational quantities---namely the  pressure $p(r)$,  energy density $\rho(r)$, and  gravitational mass $M(r)$ at a distance r from centre of star---are defined as the sums of their respective fluid components:

\begin{align}
    p(r) &= \sum_{i} p_i(r), \label{eq:p} \\
    \rho(r) &= \sum_{i} \rho_i(r), \label{eq:rho} \\
    M(r) &= \sum_{i} M_i(r). \label{eq:M}
\end{align}
where the index $i$ denotes either the DM or BM component.

 \textcolor{blue}{The expression for $F(r)$ and $Q(r)$ must be adjusted for the two-fluid scenario. To handle this, we 
 follow the derivation presented in 
 Appendix B of \citep{das2022dark}. In the two-fluid framework, the expressions for $F(r)$ and $Q(r)$ are modified as follows:
\begin{equation}
    F(r) = \frac{r - 4\pi r^3 \bigl( \sum\limits_{i} \rho_i(r) - \sum\limits_{i} p_i(r) \bigr)}{r(r - 2\sum\limits_{i}M_i(r))}
\end{equation}
\vspace{0.6cm}
 \begin{equation}
\begin{aligned}
    Q(r) &= 4\pi e^{\lambda(r)} \left[ 5\sum\limits_{i} \rho_i(r) + 9\sum\limits_{i} p_i(r) + \sum_{i} \frac{\rho_i(r) + p_i(r)}{\mathrm{d}p_i/\mathrm{d}\rho_i} \right] \\
    &\quad - \frac{6 e^{\lambda(r)}}{r^2} - \bigl[\nu'(r)\bigr]^2.
\end{aligned}
\label{eq:Q_expression}
\end{equation}}
\textcolor{blue}{Now in this two-fluid system, the quantity $y_R (\equiv y(R))$ can be determined by solving the differential equation (\ref{eq:diffeqn}), using the above  modified expressions for $\rho(r)$, $p(r)$, $M(r)$, $F(r)$, and $Q(r)$ (Eqns.(\ref{eq:p}-\ref{eq:Q_expression})) in the two-fluid framework along with TOV equation and the appropriate boundary conditions discussed in section \ref{sec:2}. The dimensionless tidal deformability then one can compute from relation $\Lambda = \frac{2}{3}k_2\beta^{-5}$. It should be noted that, in a two-fluid system,   $y(r)$ is modified because of the presence of multiple fluids inside the NS but the  equation of $y(r)$ outside the star remains unchanged (see Ref.\citep{hinderer2008tidal} for details). }

In a DANS, the calculation of the metric functions $y$ and  compactness $\beta$, and hence of the quadrupole Love number $k_2$, requires evaluation at the system's outermost boundary. Consequently, the tidal deformability is dependent on the  gravitational radius, which can differ from the visible radius $R_B$. The relevant boundary is the DM halo radius ($R = R_D$) for an extended halo configuration, and the visible radius ($R = R_B$) for a compact DM core. Furthermore, the stiffness or softness of the underlying EoS modulates the tidal response primarily through its effect on the parameter $k_2$.

\section{Result}
\label{44444}
\subsection{\textbf{Dark Matter Halo and Dark Matter Core Formation Scenarios}}
\begin{figure}[!th]
    \centering
    \begin{tabular}{cc}
        \includegraphics[scale=0.40]{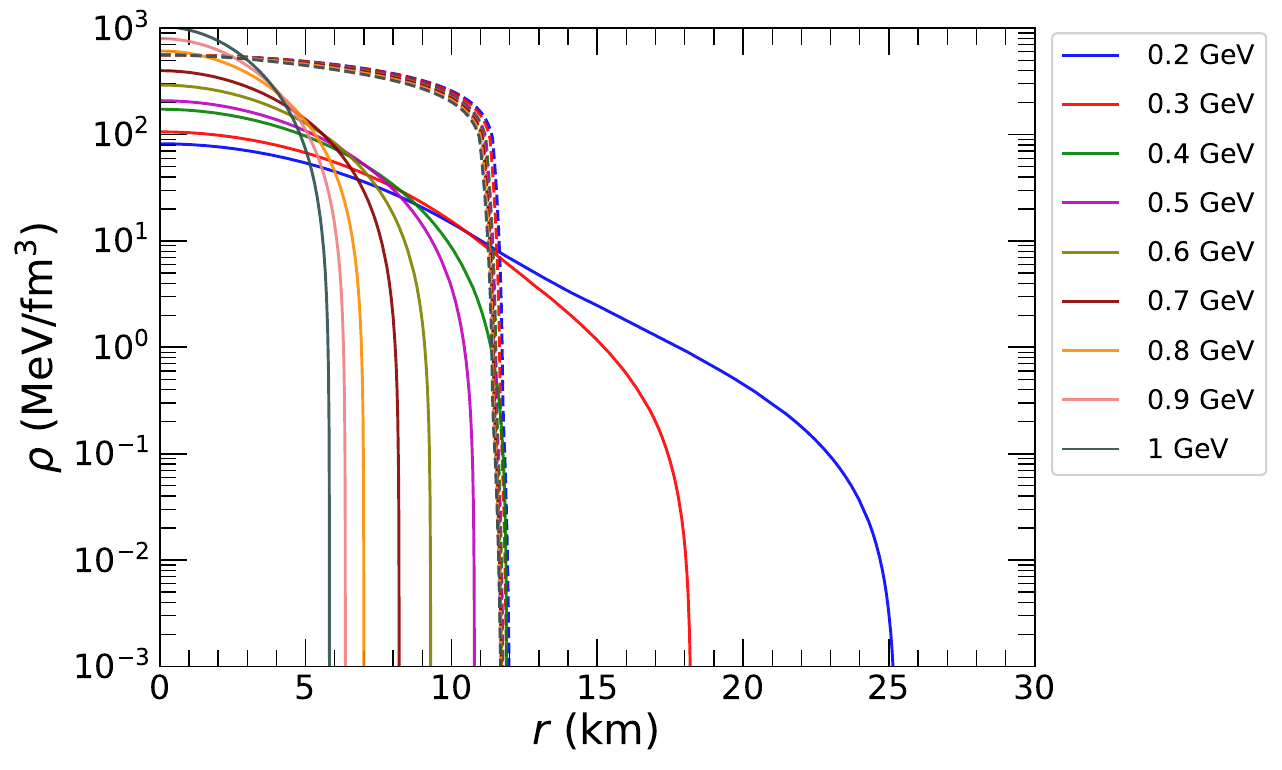} &
        \includegraphics[scale=0.40]{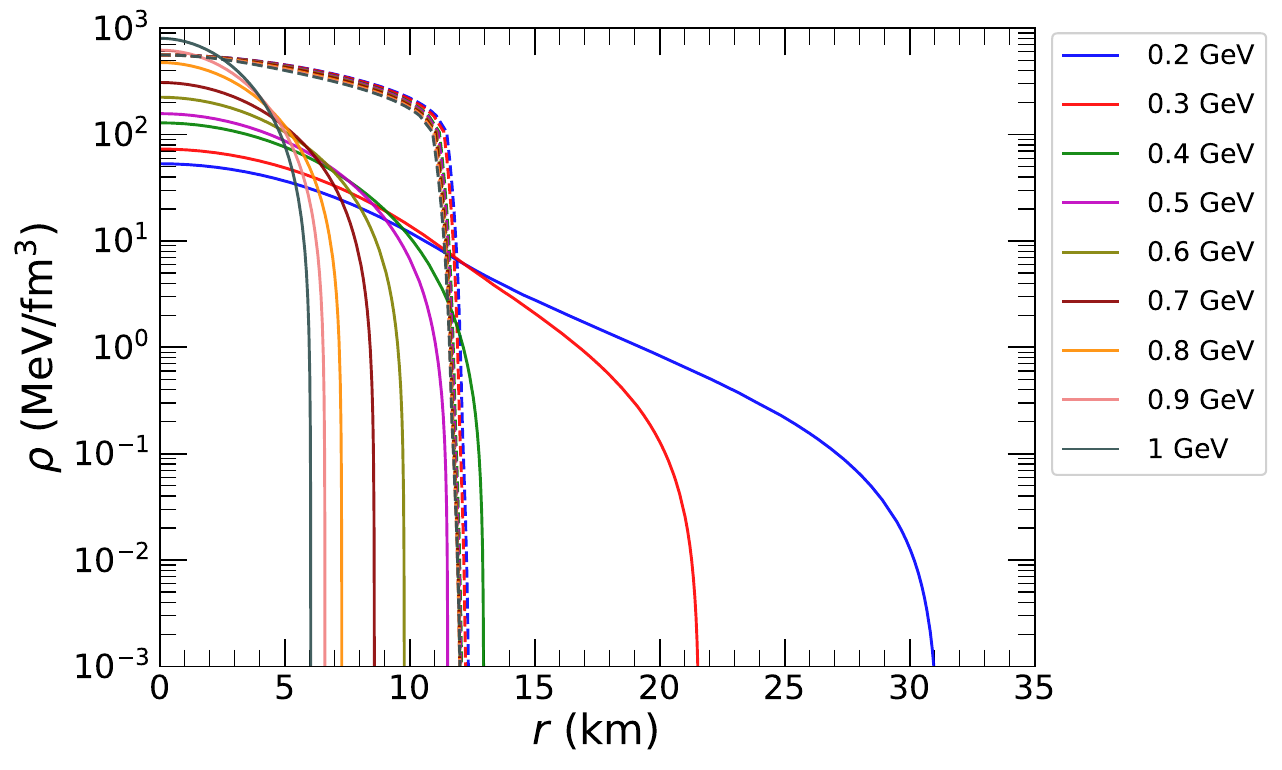}
    \end{tabular}
    \caption{Energy density profiles of DM admixed NSs for fixed DM mass fraction $f=10\%$ and different values of DM particle mass $\mu$. The DM and BM components are shown with solid and dashed lines, respectively. Results are shown for the MPA1 (left panel) and FSU2R (right panel) EoSs.}
    \label{fig:placeholder}
\end{figure}

\begin{figure}[!th]
    \centering
    \begin{tabular}{cc}
        \includegraphics[scale=0.40]{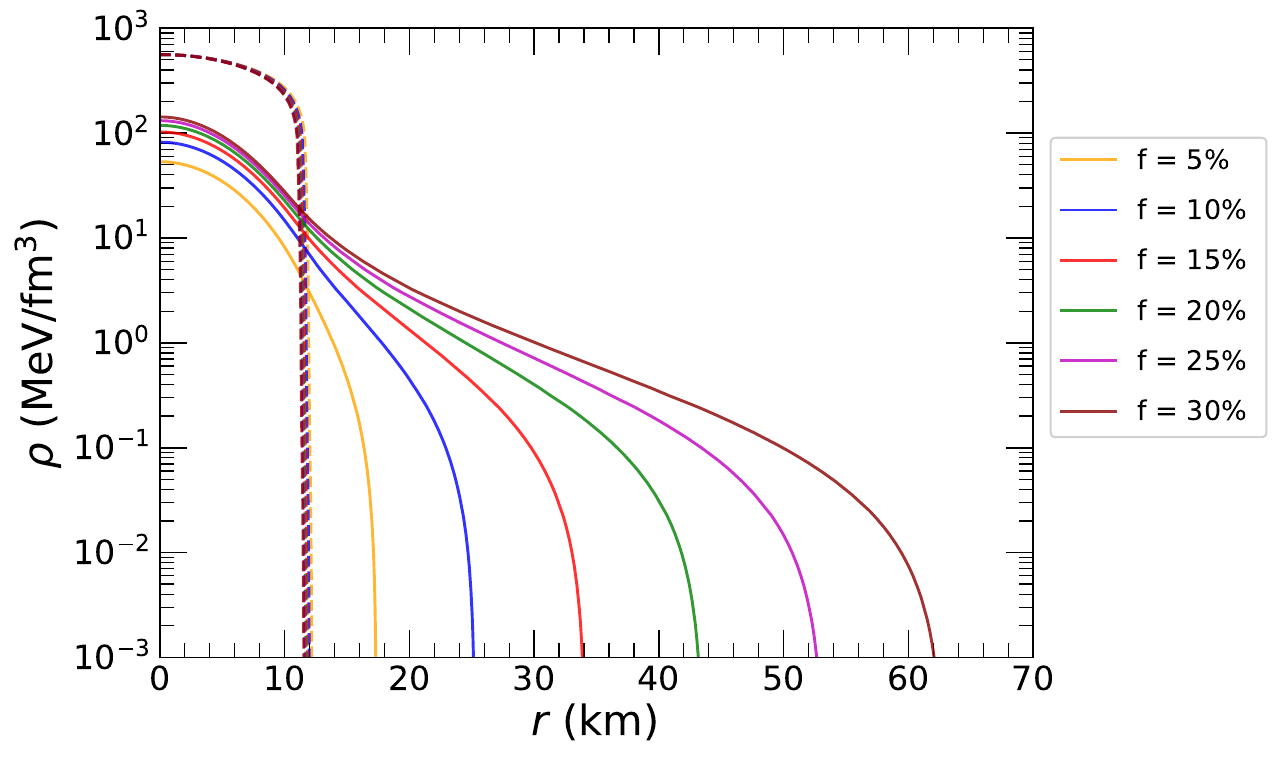} &
        \includegraphics[scale=0.40]{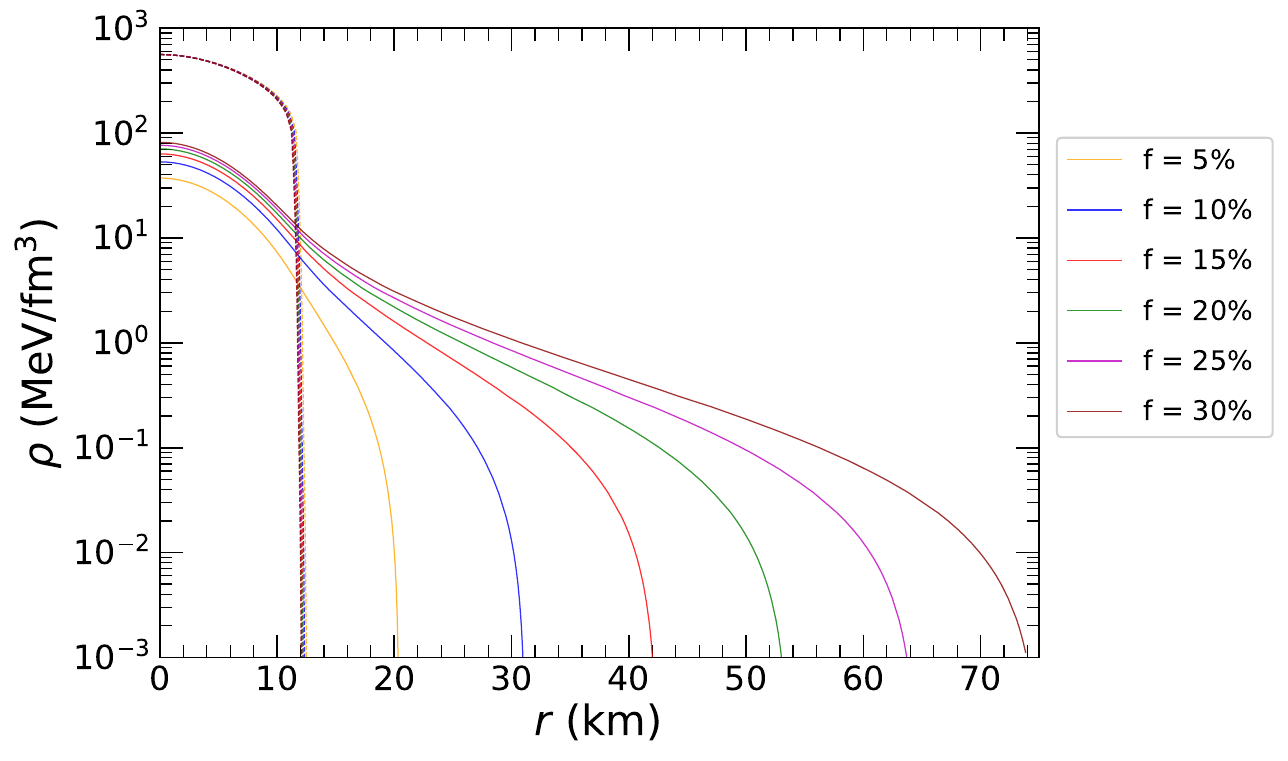}
    \end{tabular}
    \caption{Energy density profiles of DANSs for a fixed particle mass $\mu = 0.2$ GeV and varying mass fractions $f = 5\%-30\%$. Results are shown for both the MPA1 (left panel) and FSU2R (right panel) EoSs. The DM and BM components of radii are shown with solid and dashed lines respectively.}
    \label{fig:placeholdered}
\end{figure}

\begin{figure}[!th]
    \centering
    \begin{tabular}{cc}
       \includegraphics[scale=0.40]{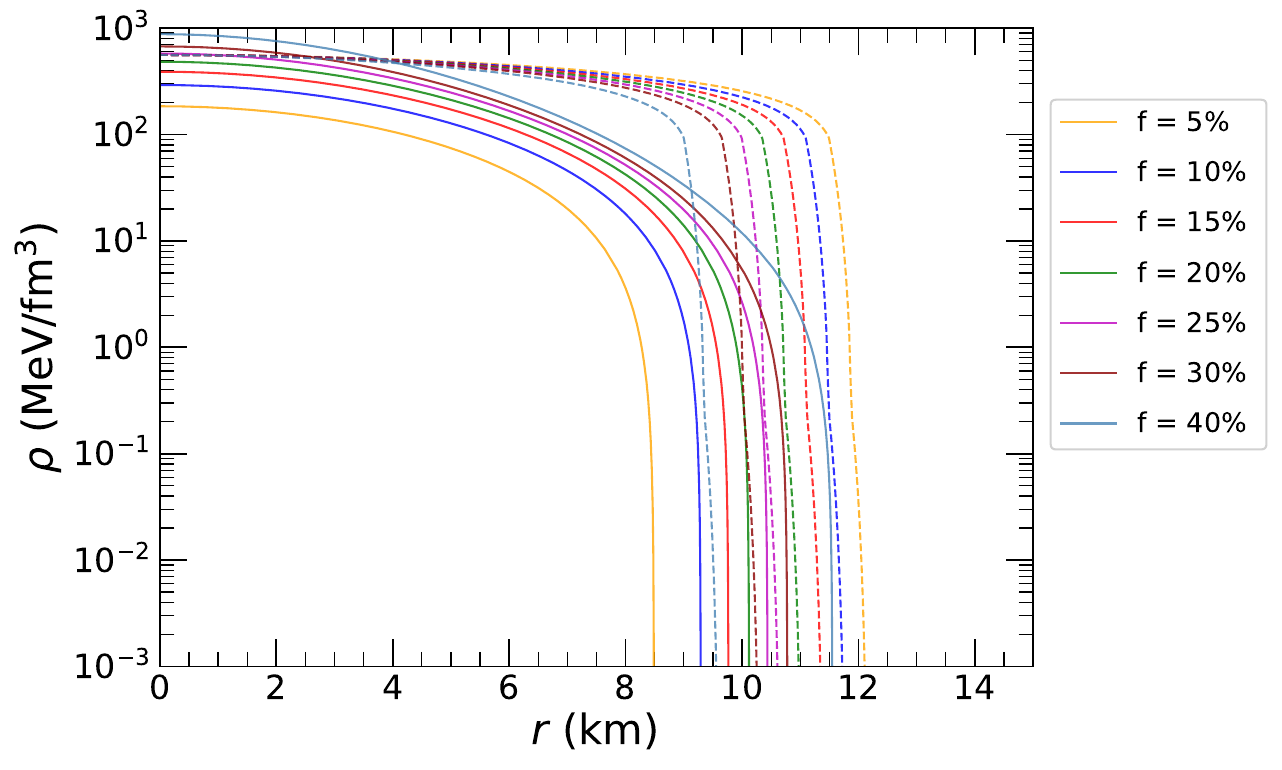} &
        \includegraphics[scale=0.40]{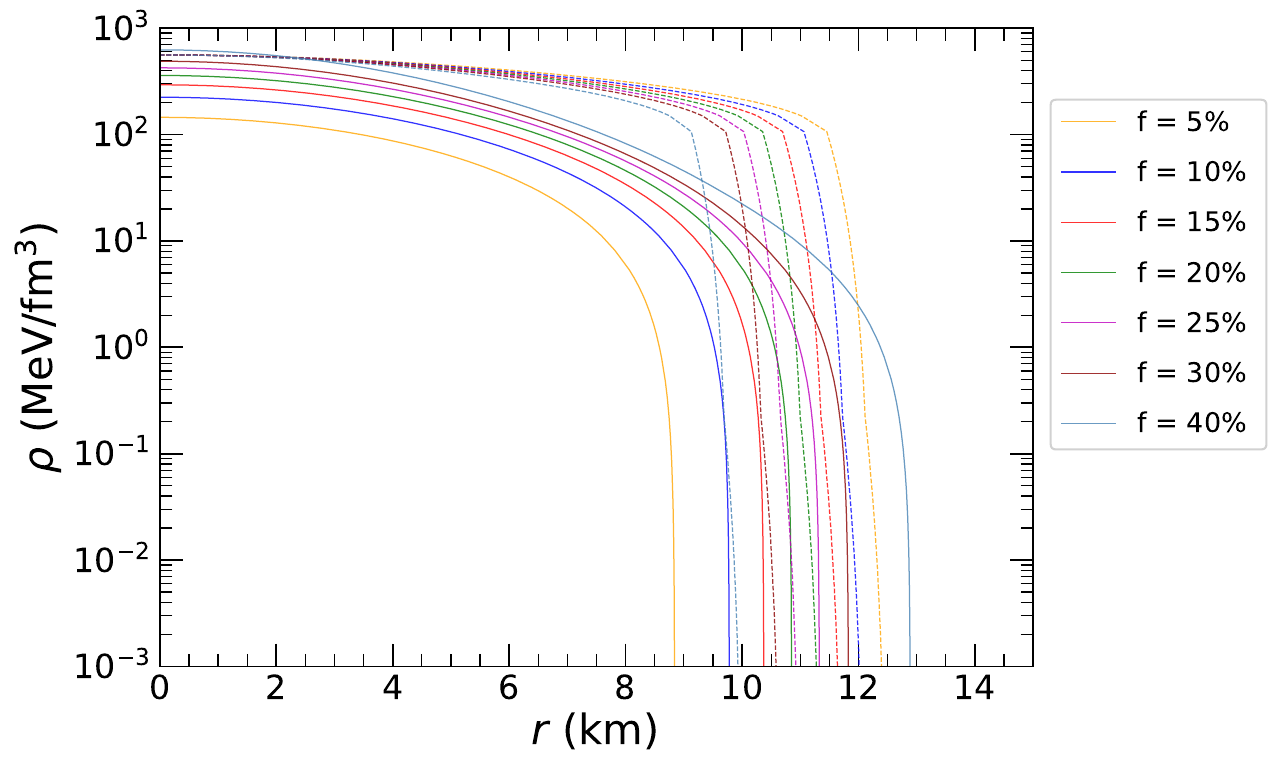}  
    \end{tabular}
    \caption{Energy density profiles of DANSs for a fixed particle mass $\mu = 0.6$ GeV and varying mass fractions $f = 5\%-40\%$. Results are shown for both the MPA1 (left panel) and FSU2R (right panel) EoSs. The DM and BM components of radii are shown with solid and dashed lines respectively.}
    \label{fig:placeholdereded}
\end{figure}

As the spatial distribution of DM depends on both the particle mass $\mu$ and mass fraction $f$,we perform an important analysis to characterize the role of these parameters. Fig.\ref{fig:placeholder} shows the variation of energy density with DM and BM radii for fixed DM fraction $f=10\%$ and different DM particle masses $\mu$. As shown in the left panel, at $\mu = 0.4$ GeV the DM radii (solid curves) and BM radii (dashed curves) coincide ($R_B \approx R_D$). Reducing $\mu$ slightly leads to DM halo formation with $R_D > R_B$, while higher particle masses produce dark matter cores where $R_B>R_D$. 
The right panel indicates that halo structures emerge for $\mu < 0.5$ GeV.
An alternative configuration is illustrated in Fig. ~\ref{fig:placeholdered}, which displays the energy density profiles for fixed mass $\mu = 0.2$ GeV and varying mass fraction $f$. It is seen that light DM particles form an extended halo surrounding the baryonic star. The halo radius is significantly larger than that of the BM component and exhibits a clear dependence on $f$. Fig.~\ref{fig:placeholdereded} illustrates the variation of the DM distribution as  $f$ is varied for $\mu = 0.6$ GeV. This figure illustrates the formation of a DM core occurs below a certain value of $f$ ($f\leq \, 25\%$ in left panel and $f\leq \, 20\%$ in right panel). On the other hand, halo configurations emerge when $f$ is sufficiently large for this fixed $\mu$ value.

Thus, the analysis reveals a systematic behavior: for a fixed value of $f$, light DM particles ($\mu < 0.5$GeV) generally produce extended halos enveloping the NS, while heavier particles primarily generate concentrated cores inside the compact star. For a fixed $\mu$, the DM core configuration typically arises at lower values of $f$, while the DM halo configuration occurs at higher values of $f$.

\subsection{\textbf{Impact of Fermionic DM on the Structural Parameters of NS}}
\label{subsec:999}
\begin{figure}[!th]
    \centering
    \begin{tabular}{cc}
       \includegraphics[scale=0.42]{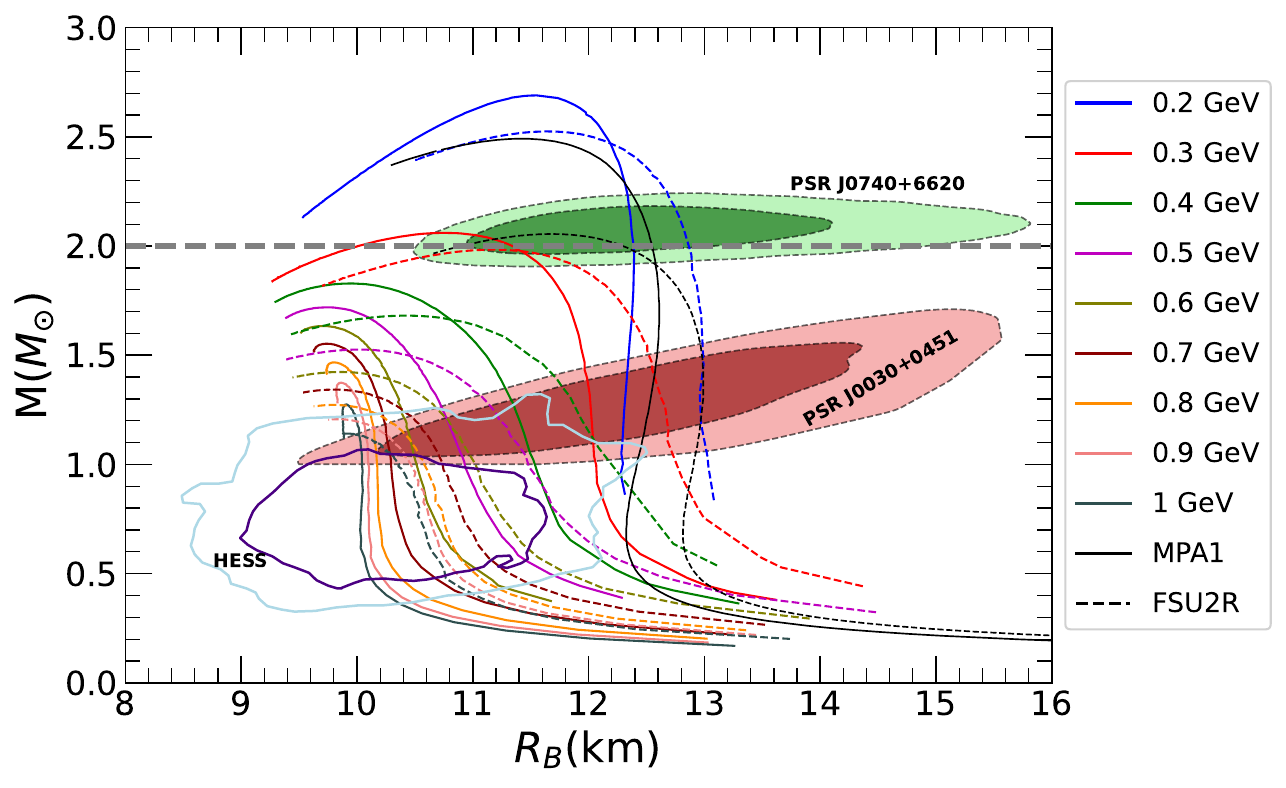}    
    \end{tabular}
    \caption{M-R curves of DANSs for various DM particle masses $\mu$ at a fixed mass fraction $f = 30\%$, with the BM described by the MPA1 (solid lines) and FSU2R (dashed lines) EoSs. The dark red (light red) and  dark green (light green) shaded areas represent the $1\sigma$ ($2\sigma$)  credible regions from NICER observations of PSR J0030+0451 and PSR J0740+6620, respectively. \textcolor{blue}{The light blue (indigo) contour lines represent the $1\sigma$ ($2\sigma$)  credible regions of HESS J1731-347 \citep{doroshenko2022strangely}}.  The horizontal gray dashed line at $2.0 M_\odot$ marks the empirical maximum mass limit for NSs.}
    \label{fig:MR1}
\end{figure}

\begin{figure}[!th]
    \centering
    \begin{tabular}{cc}
       \includegraphics[scale=0.42]{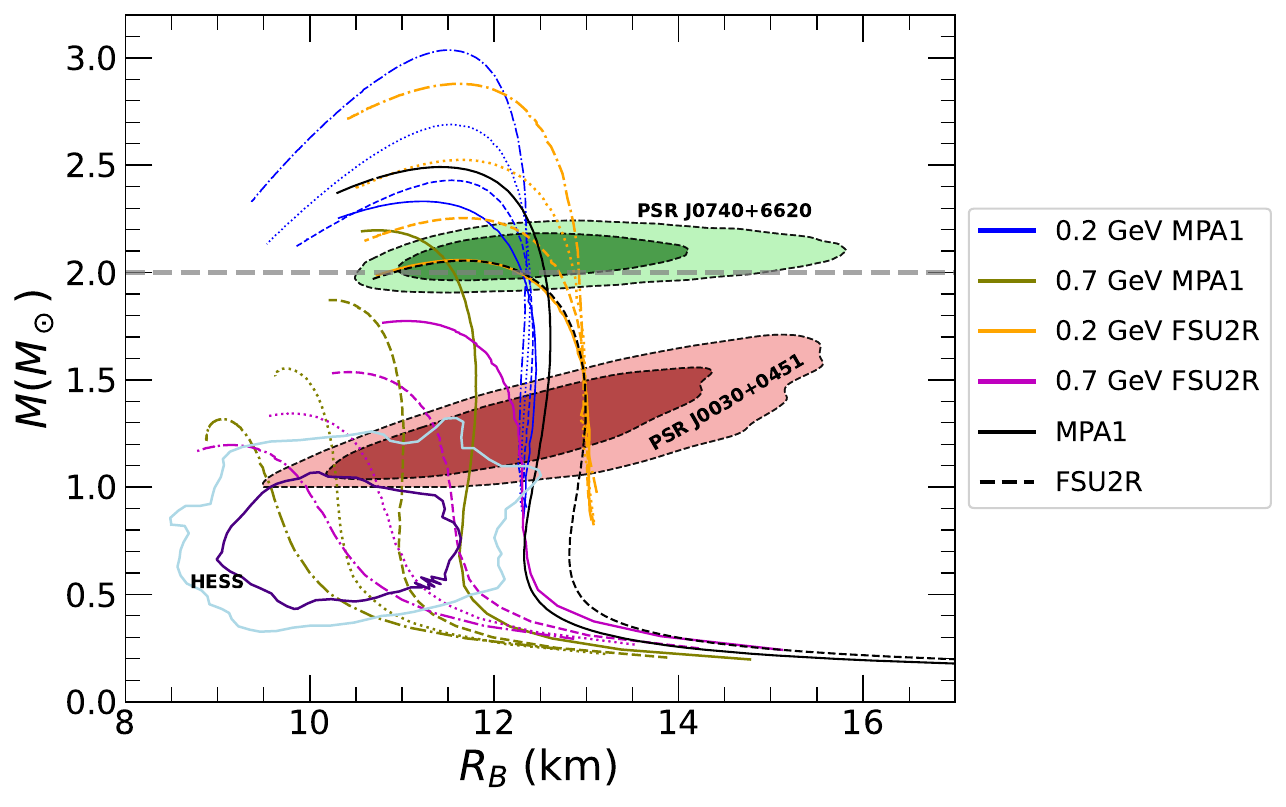}    
    \end{tabular}
    \caption{The mass–radius profiles of DANSs are shown for different DM mass fractions: $f=10\,\%$ (solid lines), $f=\,20\%$ (dashed lines),$f=30\%$ (dotted lines) and $f=40\%$ (dashed-dotted lines).Two DM configurations are considered:  DM  halo ($\mu\,=\,0.2$ GeV)  and a DM core ($\mu\,=\,0.7$) GeV, using the MPA1 and FSU2R BM EoSs.\textcolor{blue}{The dark red (light red) and  dark green (light green) shaded areas represent the $1\sigma$ ($2\sigma$)  credible regions from NICER observations of PSR J0030+0451 and PSR J0740+6620, respectively. \textcolor{blue}{The light blue (indigo) contour lines represent the $1\sigma$ ($2\sigma$)  credible regions of HESS J1731-347 \citep{doroshenko2022strangely}}.  The horizontal gray dashed line at $2.0 M_\odot$ marks the empirical maximum mass limit for NSs.}}
    \label{fig:MR2}
\end{figure}

In this section, we investigate the influence of fermionic DM  on NSs by analyzing the corresponding mass-radius (M-R) relations. Here we have considered visible radius $R_B$ rather than outermost radius $R$. This analysis employs two distinct BM EoSs —MPA1 and FSU2R—to describe the nuclear matter within the NS. Our theoretical M-R profiles are subsequently compared with the latest observational constraints from the Neutron Star Interior Composition Explorer (NICER) \citep{miller2019psr,riley2019nicer,miller2021radius,riley2021nicer}. Figures ~\ref{fig:MR1} and ~\ref{fig:MR2} display the NICER credible regions for PSR J0030+0451 
(red) and PSR J0740+6620 (green), with the darker and lighter shades representing the $68\%$ ($1\sigma$) and $95\%$ ($2\sigma$) confidence levels, respectively. \textcolor{blue}{The $1\sigma$ ($2\sigma$) credible regions for the light compact object $HESS\,J1731-347$ ($M\,\approx\,0.77M_{\odot}$, $R\,\approx\,10.4$ km) \citep{doroshenko2022strangely} are indicated by light-blue (indigo) contour lines}. NICER results favor NS models with relatively large radii, typically  $R_B\gtrsim\,11$ km. Figure ~\ref{fig:MR1} presents the M-R relations for DANSs with a fixed $f$ of $30\%$ across a range of $\mu$ from $0.2$ to $1\ \mathrm{GeV}$. An increase in $\mu$ is shown to lead to a decrease in the $M_{max}$. For massive fermions ($\mu\geq 0.8$ GeV) which correspond to the DM core configuration, the resulting M-R profiles of DANSs with the MPA1 EoS (solid lines) fall outside NICER's $68\% (1\sigma)$ credible regions for PSR J0030+0451 (dark red region). In contrast, lighter DM particle masses ($\mu <0.4\, \mathrm{GeV}$), which give rise to a DM halo rather than a core, produce M-R sequences that remain compatible with the credible region defined by PSR J0740+6620. The M-R curves corresponding to DM core configuration ($\mu \geq 0.4\ \mathrm{GeV}$) are inconsistent with the credible region associated with PSR J0740+6620. \textcolor{blue}{We further observe that heavy DM particles, associated with DM core configurations, generate M–R curves that fall well within the HESS region. On the other hand, the M–R curves associated with the DM halo configuration fall outside the HESS region. Thus, one can conclude that only DM core models with sufficiently massive fermions are capable of reproducing the HESS like light compact object.}

Figure ~\ref{fig:MR2} illustrates the influence of $f$ on the M-R relations of mixed compact objects with DM core and halo configurations. Two representative DM particle masses are considered: $\mu = 0.2\ \mathrm{GeV}$, corresponding to a DM halo formation, and $\mu = 0.7\ \mathrm{GeV}$, corresponding to a DM core formation, distinguished by distinct colors for both the MPA1 and FSU2R BM EoSs. Different line styles denote varying $f$: solid lines for $f = 10\%$, dashed lines for $f = 20\%$, dotted lines for $f = 30\%$, and dashed-dotted lines for $f = 40\%$.
For the core configuration ( for $\mu = 0.7\ \mathrm{GeV}$), increasing the $f$ value from $10\%$ to $40\%$ reduces both the $M_{max}$ and $R_B$ of the mixed object, shifting the M-R curves away from the observationally allowed parameter spaces. In contrast, for the  DM halo configuration ( for $\mu = 0.2\ \mathrm{GeV}$),$M_{max}$ increases with the  $f$ (from $10\%$ to $40\%$) while the baryonic radius $R_B$ stays approximately constant. \textcolor{blue}{For the DM core configuration produced by massive fermions, the curves associated with larger DM fractions fall comfortably within the HESS $1\sigma$ region.}

Therefore, our analysis in this subsection demonstrates that fermionic DM with lower masses (i.e., DM halo configurations) yields DANSs that exhibit greater compatibility with the latest NICER constraints across the entire range of $f$ considered here. On the other hand, a DM core configuration with a low value of $f$ is strongly favored by the NICER constraints for both PSR J0030+0451 and PSR J0740+6620. \textcolor{blue}{Also the ultra-light compact object HESS J1731-347  can be explained by a fermionic DANS with a heavy DM core and large DM fraction.}

Furthermore, our results show that DANSs with an ideal Fermi gas DM component and a baryonic MPA1 or FSU2R EoS can account for several exotic compact objects. These range from the very massive ($2.6M_{\odot}$) secondary component of GW190814 binary merger event \citep{abbott2020gw190814} to the very light candidate HESS J1731-347 \citep{doroshenko2022strangely} with a mass of approximately $0.77 M_\odot$.

\begin{figure}[!th]
    \centering
    \begin{tabular}{cc}
       \includegraphics[scale=0.40]{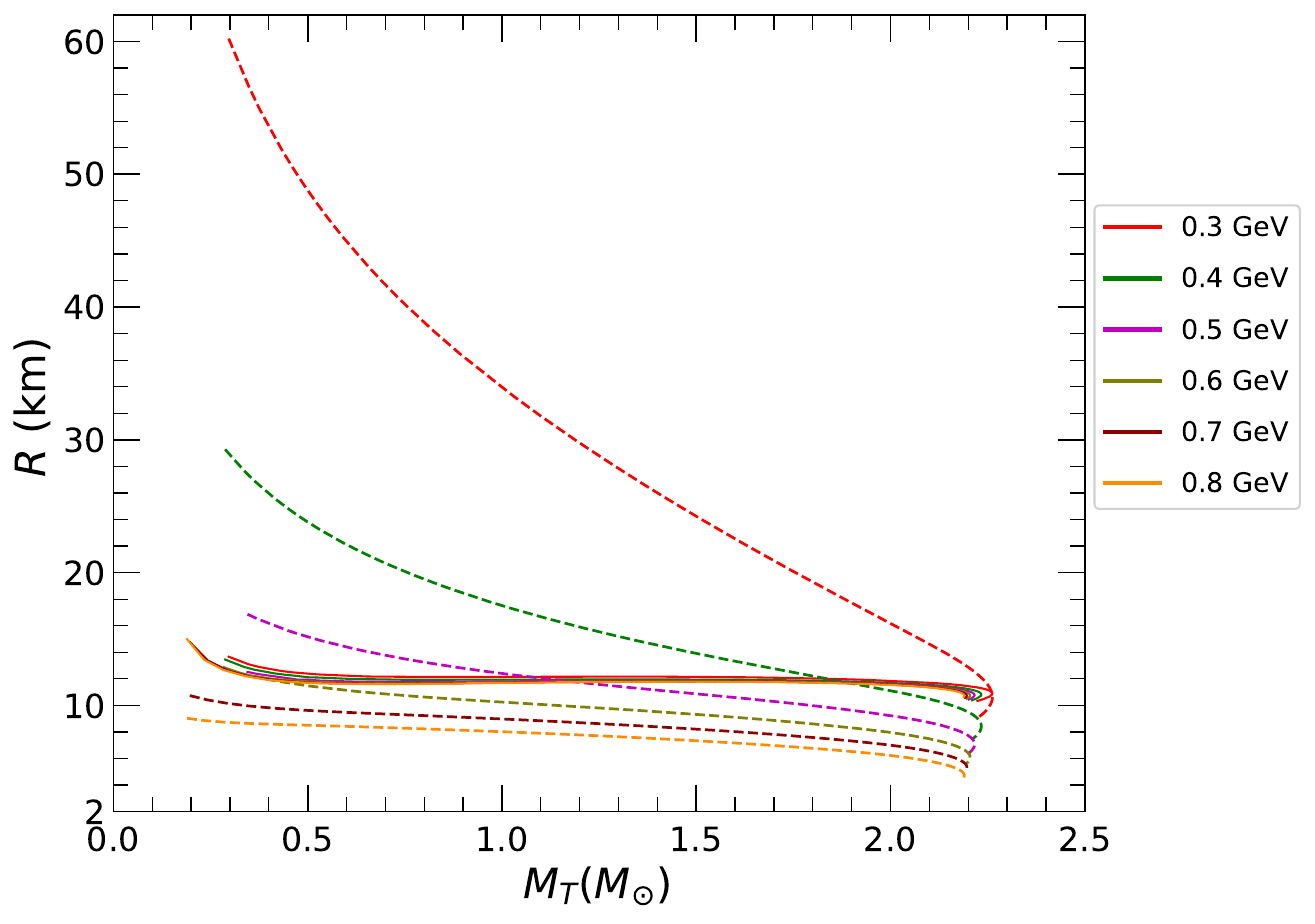} &
        \includegraphics[scale=0.39]{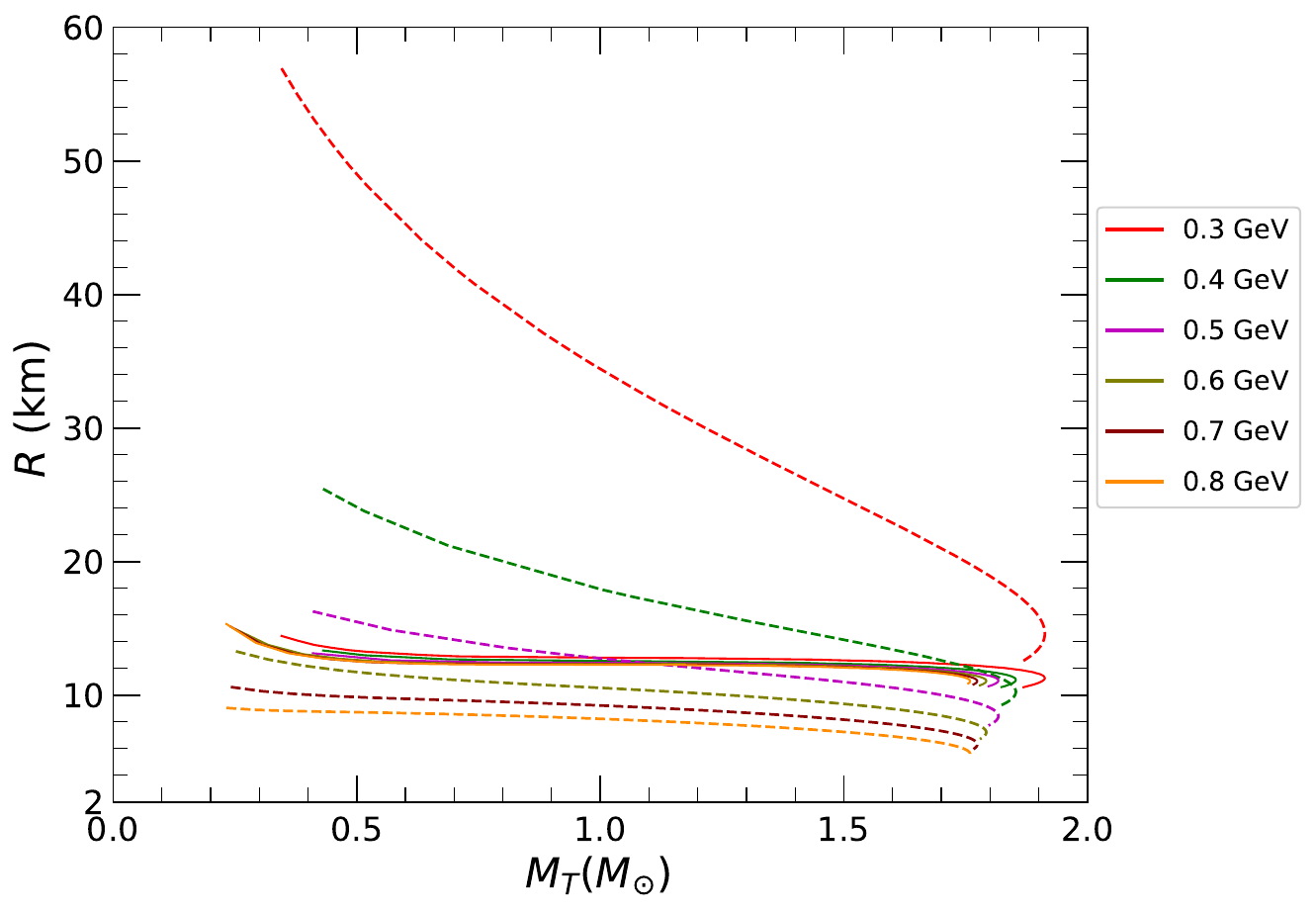}  
    \end{tabular}
    \caption{Radii of the DM ($R_D$) and BM ($R_B$) components as a function of the total gravitational mass $M_T$, calculated for a DM fraction \( f = 10\% \) and various fermion masses \( \mu \). Results for the MPA1 and FSU2R BM EoSs are shown in the left and right panels respectively. The solid and dashed curves represent the  \( R_B \) and  \( R_D \) respectively. Configurations with a DM halo is indicated by the red (\( \mu = 0.3\ \text{GeV} \))  curve. The green curve (\( \mu = 0.4\ \text{GeV} \)) marks the transition from a DM core to a halo. Curves for \( \mu \geq 0.5\ \text{GeV} \) correspond to pure DM core configurations.}
    \label{fig:MR3}
\end{figure}

Fig. ~\ref{fig:MR3} presents the radii of the BM ($R_B$) and DM  ($R_D$) components separately as functions of the total gravitational mass $M_T$, for a fixed DM mass fraction $f=10\%$ and various values of the fermion mass $\mu$ for both MPA1 (left panel) and FSU2R (right panel) BM EoSs. For smaller values of $\mu$, $R_D$ is larger than $R_B$ across entire mass range, indicating an extended DM halo configuration. As $\mu$ increases, a structural transition becomes evident. For example, for $\mu=0.4$ GeV (green dashed curve), $R_D>R_B$ only up to an intermediate range of $M_T$; then $R_B$ becomes larger than $R_D$. This behavior signals a transition from a DM halo to a DM core configuration for $f=10\%$. 
\begin{figure}[!th]
    \centering
    \begin{tabular}{cc}
       \includegraphics[scale=0.40]{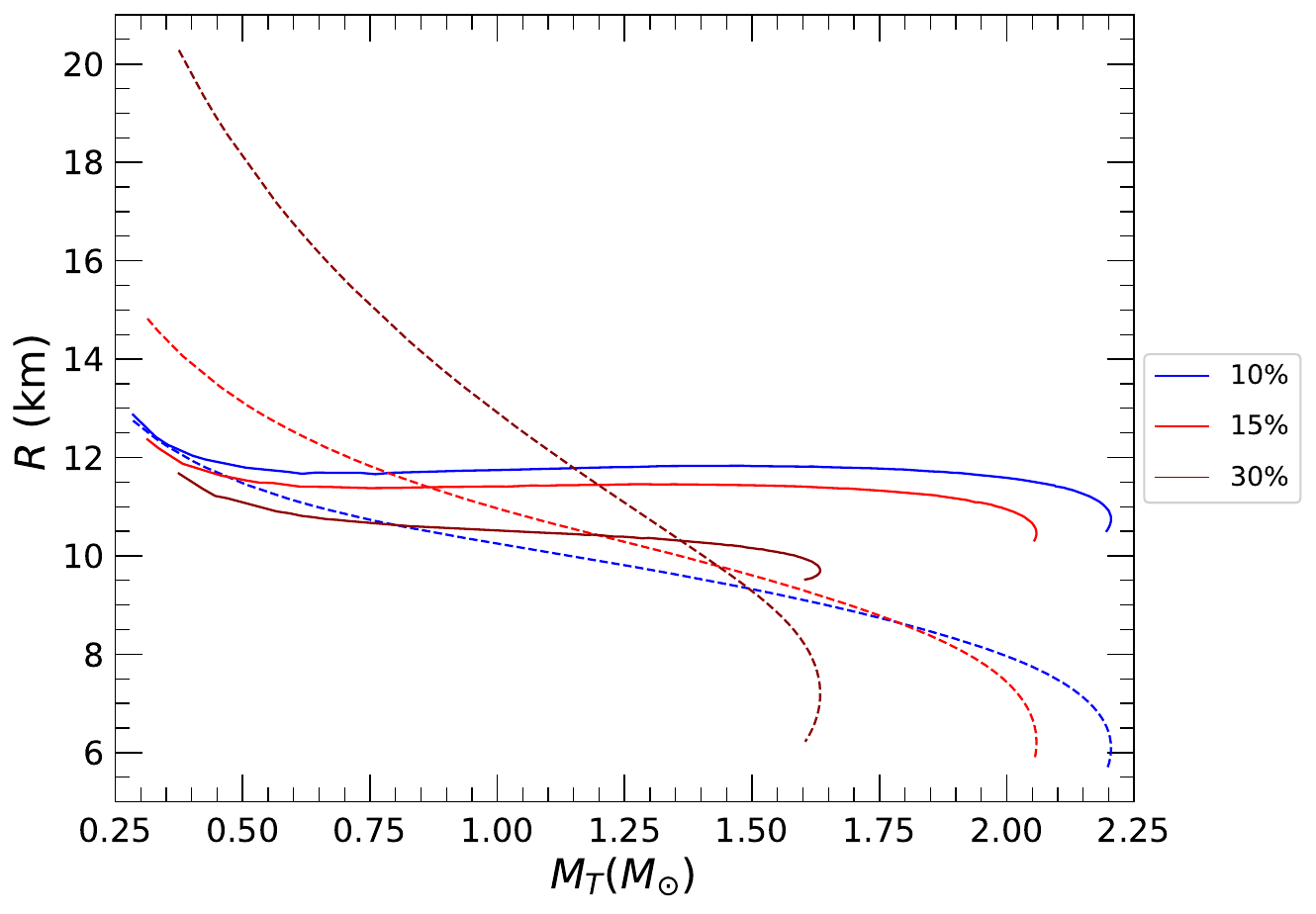 }  &
        \includegraphics[scale=0.39]{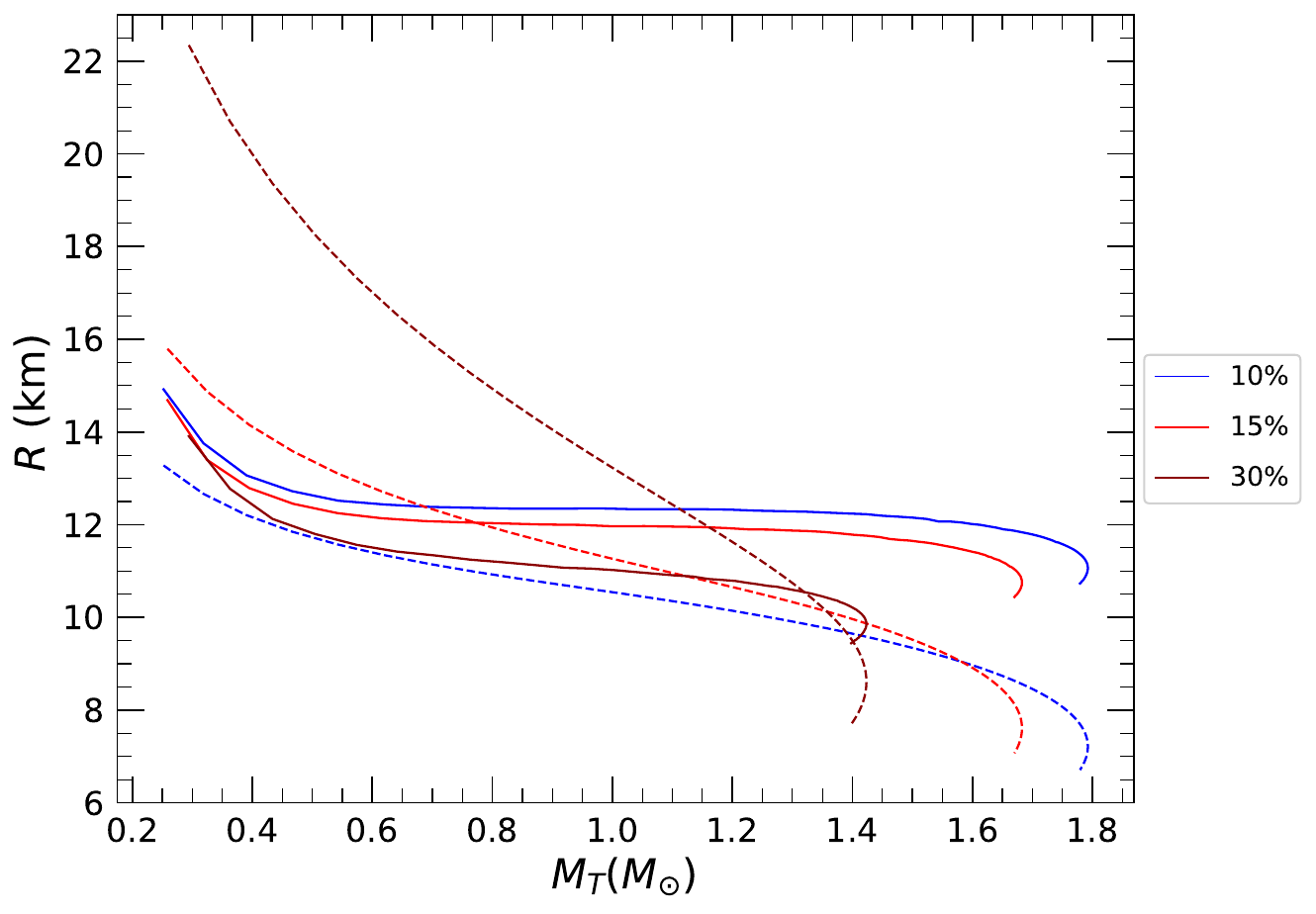}  
    \end{tabular}
    \caption{Radii of the DM ($R_D$) and BM ($R_B$) components as a function of the total gravitational mass $M_T$, calculated for various $f$ and  for a fixed fermion mass \( \mu\,=\,0.6 \) GeV. Results for the MPA1 and FSU2R BM EoSs are shown in the left and right panels, respectively. The solid and dashed curves represent the  \( R_B \) and  \( R_D \) respectively.}
    \label{fig:MR new}
\end{figure}
For massive fermions ($\mu\geq 0.5$ GeV), the condition $R_B>R_D$ holds across the astrophysically relevant range of mass of NS, corresponding to a DM core structure. Fig.~\ref{fig:MR new}
shows a similar transition for $\mu = 0.6$ GeV and various values of $f$. For lower values of $f$, a DM core configuration ($R_B>R_D$) persists across the entire mass range. As $f$ increases, a transition from a DM halo configuration ($R_D>R_B$) to a DM core configuration  occurs at a certain value of $M_T$.

\begin{figure}[!th]
    \centering
    \begin{tabular}{cc}
       \includegraphics[scale=0.40]{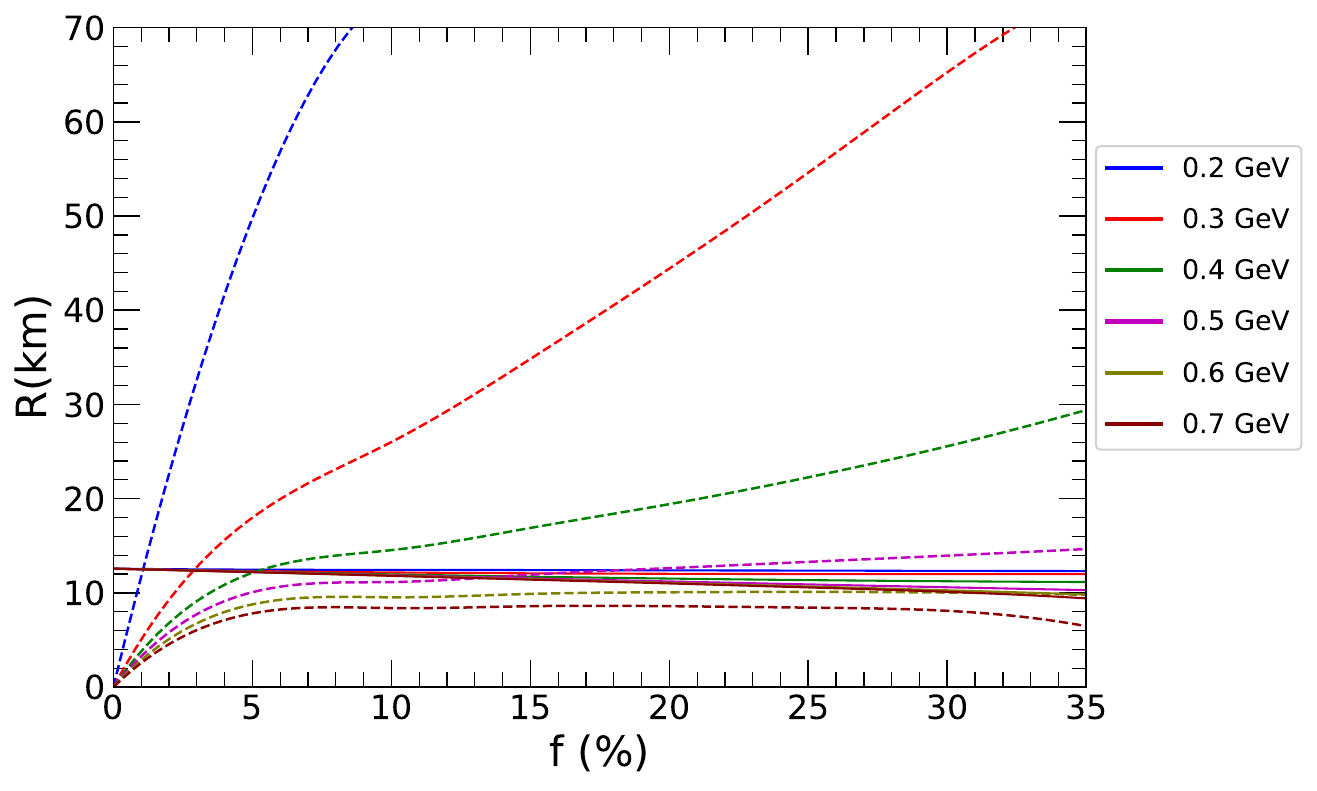}  &
        \includegraphics[scale=0.39]{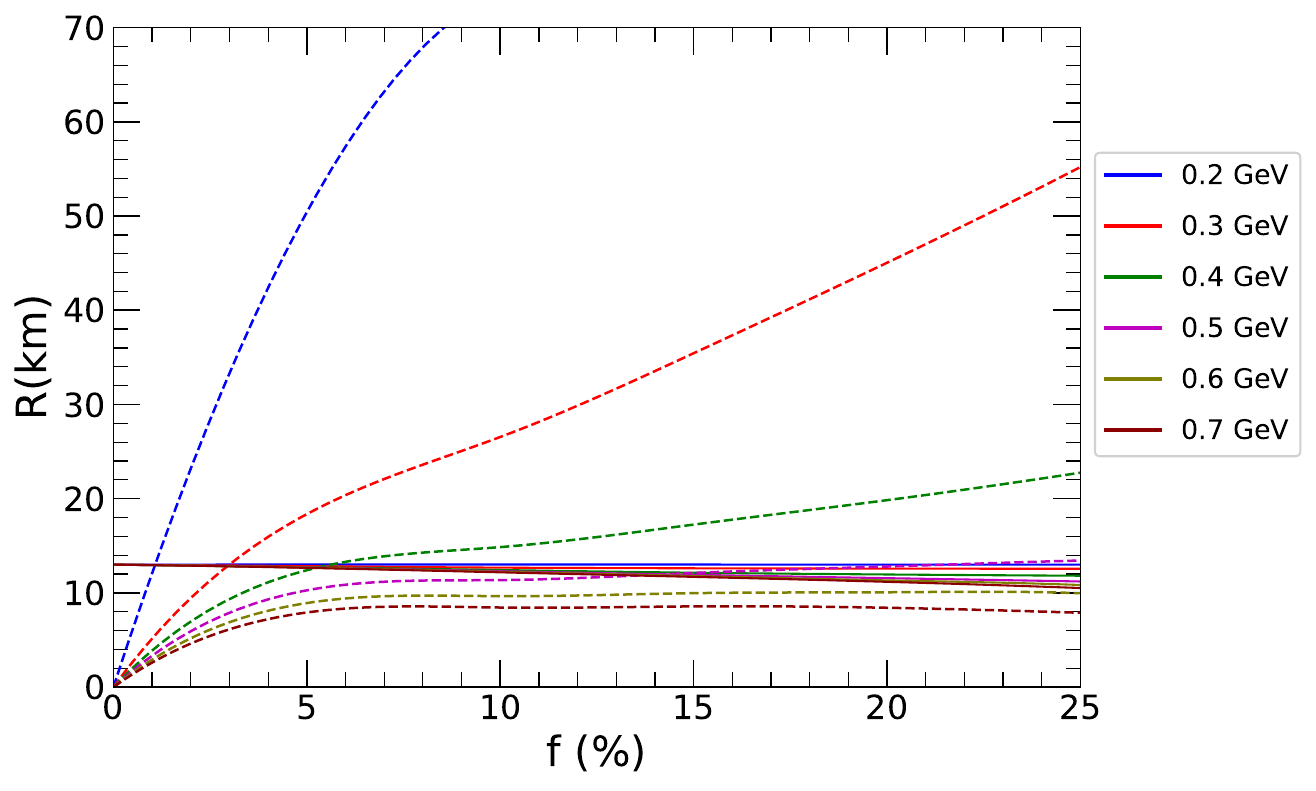}  
    \end{tabular}
    \caption{Left panel: Radii of the BM ($R_B$) ,  and DM ($R_D$) components versus  \(f\) for  DANSs having total mass \(M_T=1.4 M_\odot\) , using the MPA1 BM EoS. Results are shown for various values of $\mu$. \(R_B\)  and \(R_D\) are shown by solid and dashed lines respectively . Right panel: Same as the left panel, but for the FSU2R BM EoS.}
    \label{fig:MR4}
\end{figure}
\begin{figure}[!th]
    \centering
    \begin{tabular}{cc}
       \includegraphics[scale=0.40]{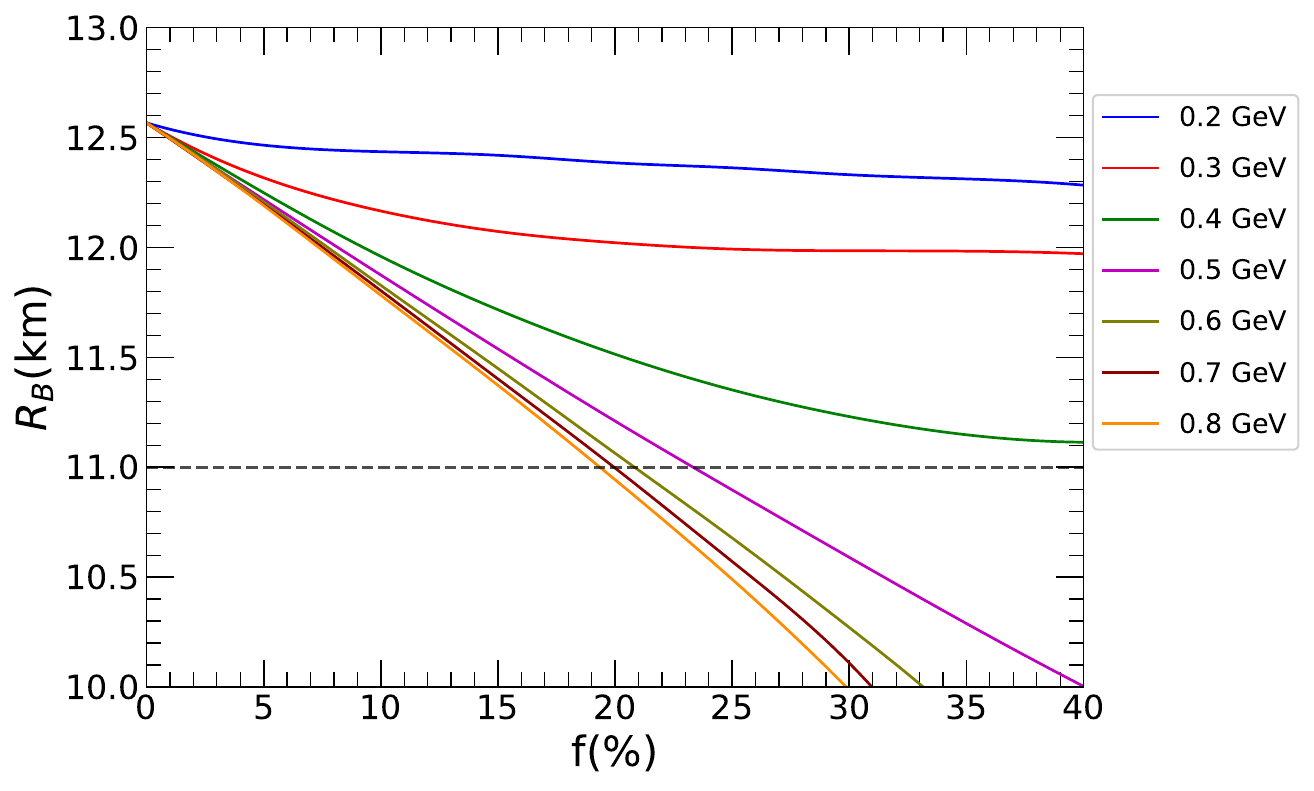} &
        \includegraphics[scale=0.39]{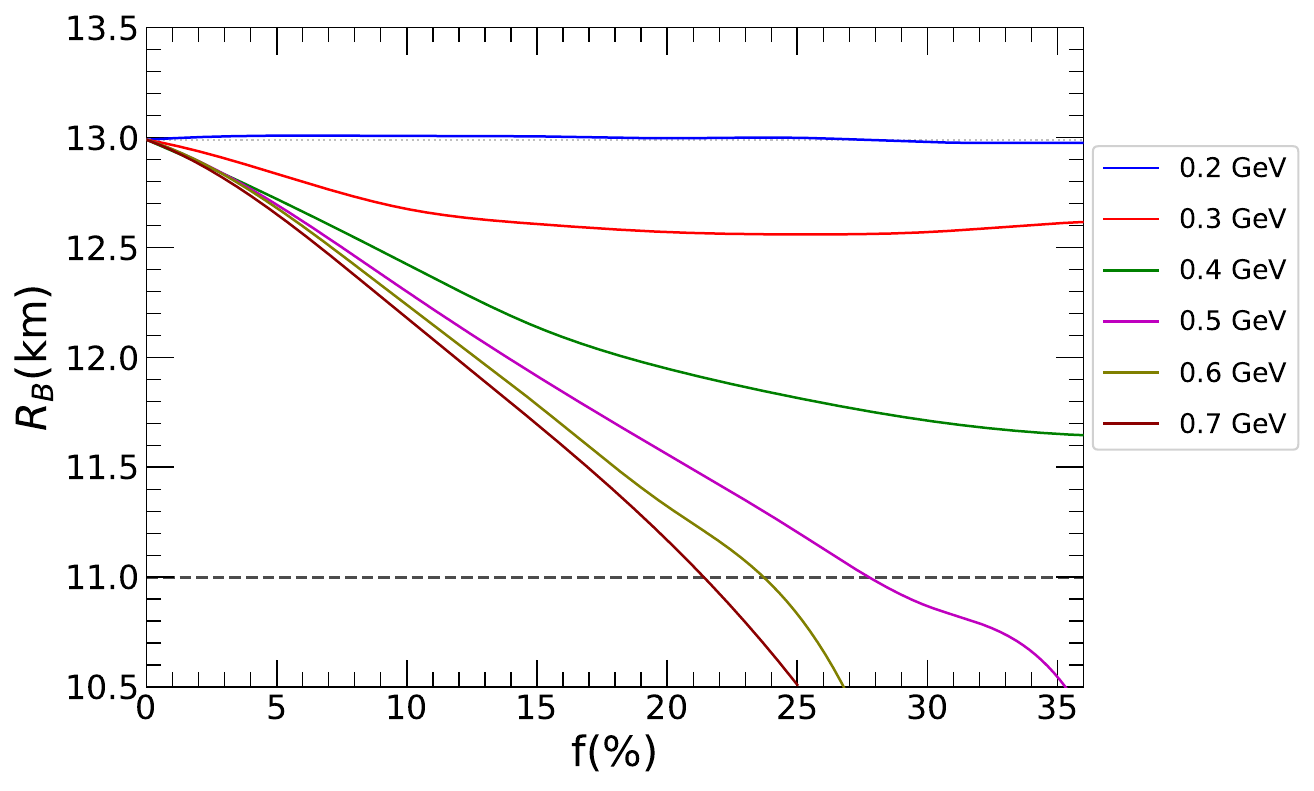}  
    \end{tabular}
    \caption{Left panel: BM radius \(R_B\) as a function of  \(f\) for DANSs having total mass \(M_T=1.4 M_\odot\), calculated using the MPA1 BM EoS. Results are shown for different values of $\mu$. The black dashed line indicates the lower observational constraint for \(R_{1.4}\). Right panel: Same as the left panel, but calculated using the FSU2R BM EoS.}
    \label{fig:MR5}
\end{figure}
\begin{figure}[!th]
    \centering
    \begin{tabular}{cc}
       \includegraphics[scale=0.40]{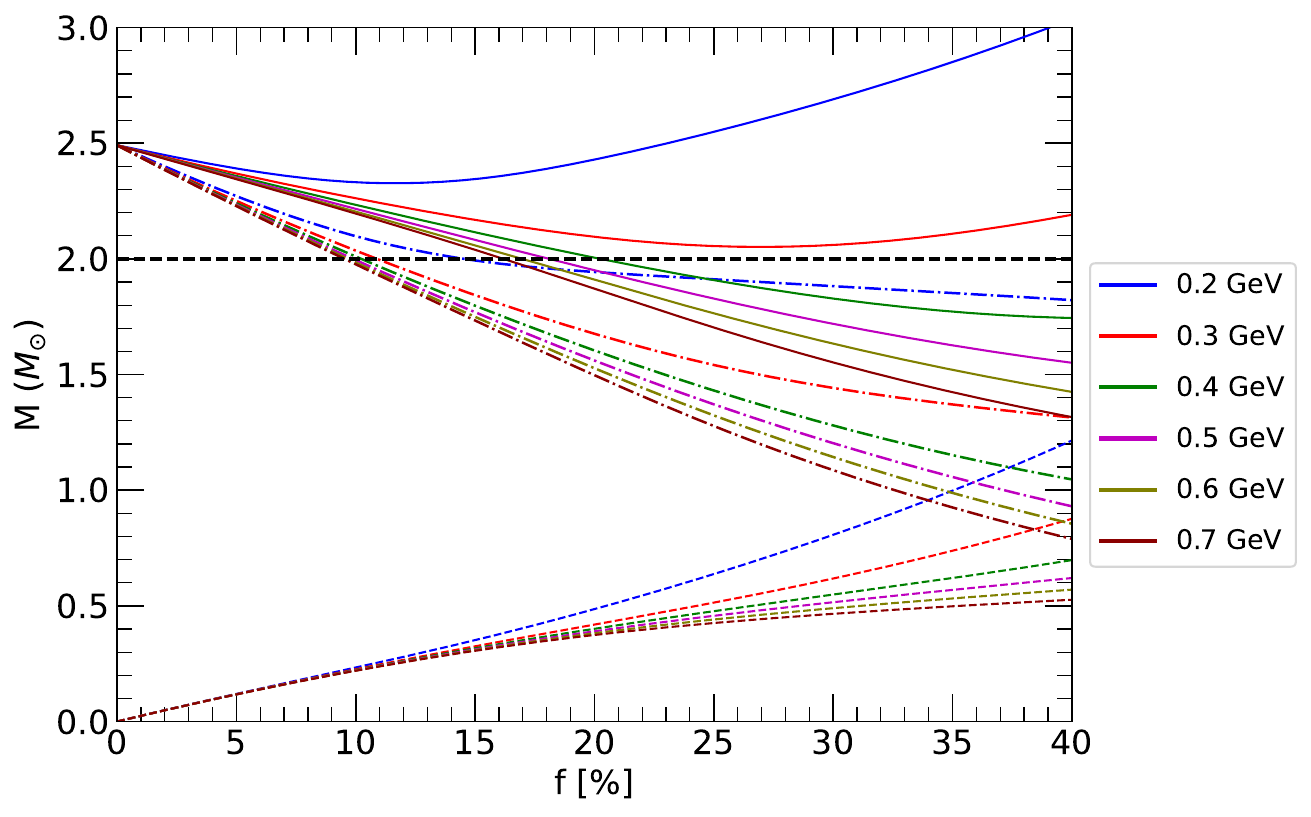} &
        \includegraphics[scale=0.40]{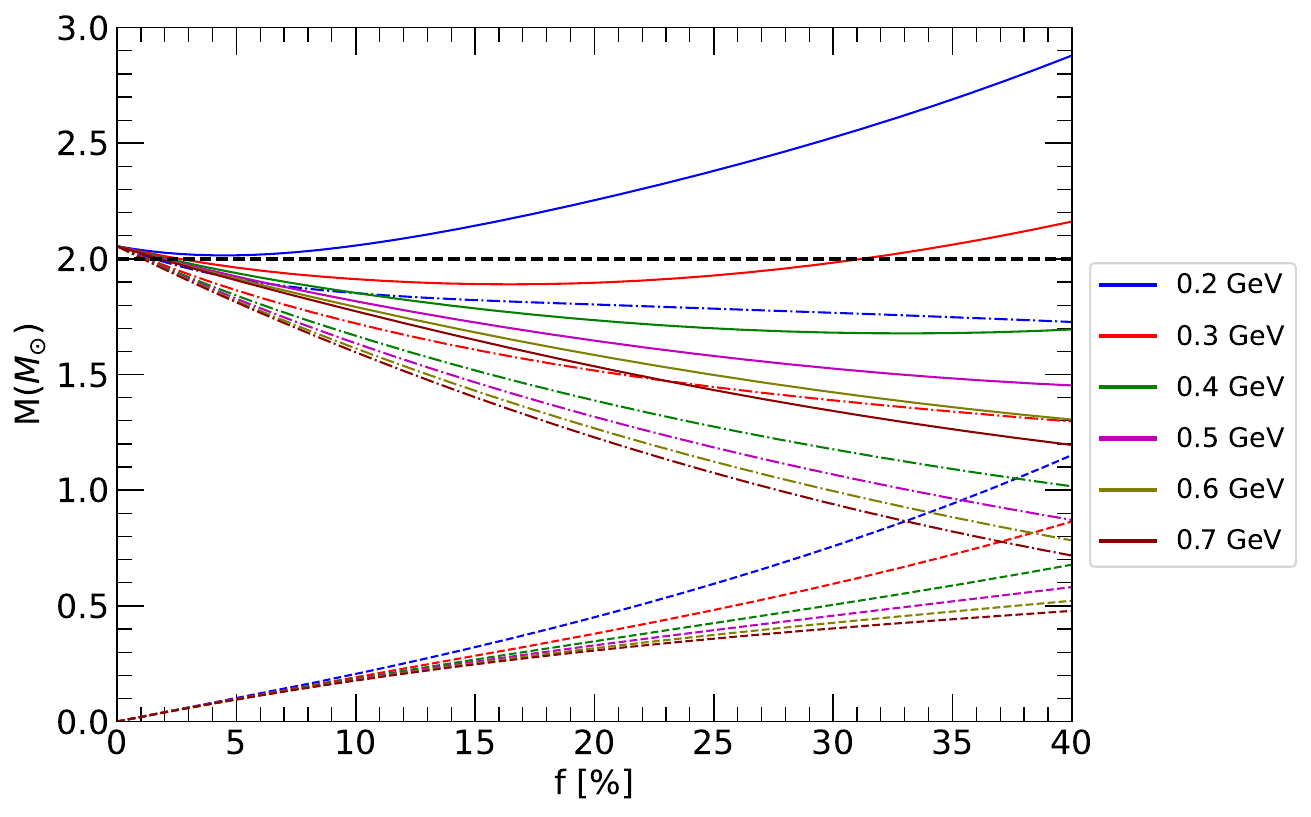}  
    \end{tabular}
    \caption{Left panel : The solid, dashed, and dash-dotted curves depict the maximum total gravitational mass ($M_{Tmax}$), the DM mass ($M_D$), and the BM mass ($M_B$), respectively for MPA1 BM EoS. The $2 M_{\odot}$ limit is marked by a horizontal black dashed line. Right panel: Same as the left panel but for the FSU2R BM EoS.}
    \label{fig:MR6}
\end{figure}
In Fig.~\ref{fig:MR4}, \(R_B\), (shown by solid lines) and  \(R_D\), ( shown by dashed lines) are shown as functions of \(f\) for different values of $\mu$, considering DANSs with  \(M_T=1.4 M_\odot\). For all considered $\mu$ values, \(R_B\) remains approximately constant at \(\approx 13\ \text{km}\), while \(R_D\) increases with \(f\). Lighter DM particles exhibit steeper growth of \(R_D\) at low fractions compared to heavier ones. The transition to a DM halo configuration appears when \(R_D\) becomes comparable to \(R_B\) (\(R_D \approx R_B\)).For massive fermions, DM core $(R_B > R_D)$ is formed inside NSs across the entire range of $f$. We also see that even for lighter fermions with mass $\mu \lesssim 0.4\ \text{GeV}$, a DM core configuration may still form, but only at very low values of $f$.

In Fig.~\ref{fig:MR5},  \( R_B \) is shown to be a decreasing function of  \( f \). This reduction in \( R_B \) is more pronounced for massive fermions, whereas for light fermions the variation gradually slows, approaching a constant value at larger \( f \). For massive fermions, an upper limit on $f$ can be established to ensure that $R_B$ satisfies  given observational constraint on radius  (\( R_B \geq 11\ \text{km} \)). In contrast, for sufficiently light fermions, \( R_B \) remains above this radius limit across the entire range of $f$ considered thereby satisfying the observational constraints on radius. It follows that increasing \( f \) for massive fermions leads to values of \( R_B \) below \( 11\ \text{km} \), which is disfavoured by the lower observational bound on the radius of a \( 1.4\,M_{\odot} \) NS. For our considered range of $\mu$ values, $R_B$ remain above \( 11\ \text{km} \) for  \( f \lesssim 20\% \) (for the MPA1 BM EoS in the left panel) and \( f \lesssim 21\% \) (for the FSU2R BM EoS in the right panel).

Figure~\ref{fig:MR6} shows how the DM (dashed curves) and BM (dot-dashed curves) components contribute to the maximum total gravitational mass $M_{Tmax}$(solid curves) in DANSs. We observe an inverse relationship: $M_B$ decreases with the DM fraction $f$, while $M_D$ increases. Notably, the rates at which $M_B$ and $M_D$ changes, differ considerably between systems with a DM halo (e.g., $\mu = 0.2$~\text{GeV}) and those with a DM core (e.g., $\mu = 0.7$~\text{GeV}). For lighter fermions, $M_{Tmax}$ (solid curves) begins to increase at higher values of $f$. As $\mu$ increases, this growth becomes progressively slower due to the transition from a DM halo to a DM core configuration.

\subsection{\textbf{Impact of Fermionic DM on the Tidal Parameters of NS}}
\begin{figure}[!th]
    \centering
    \begin{tabular}{cc}
       \includegraphics[scale=0.39]{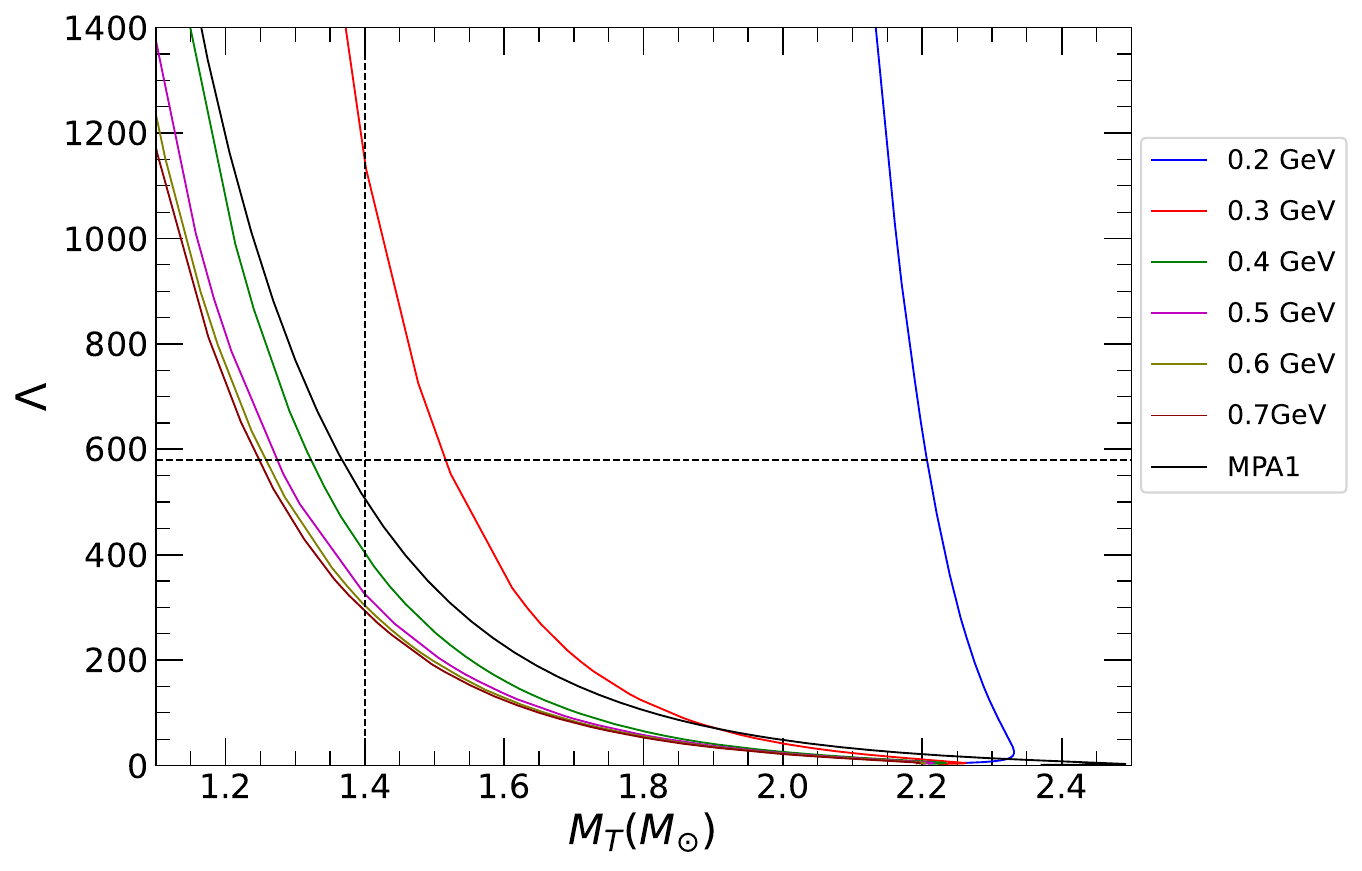}  &
        \includegraphics[scale=0.38]{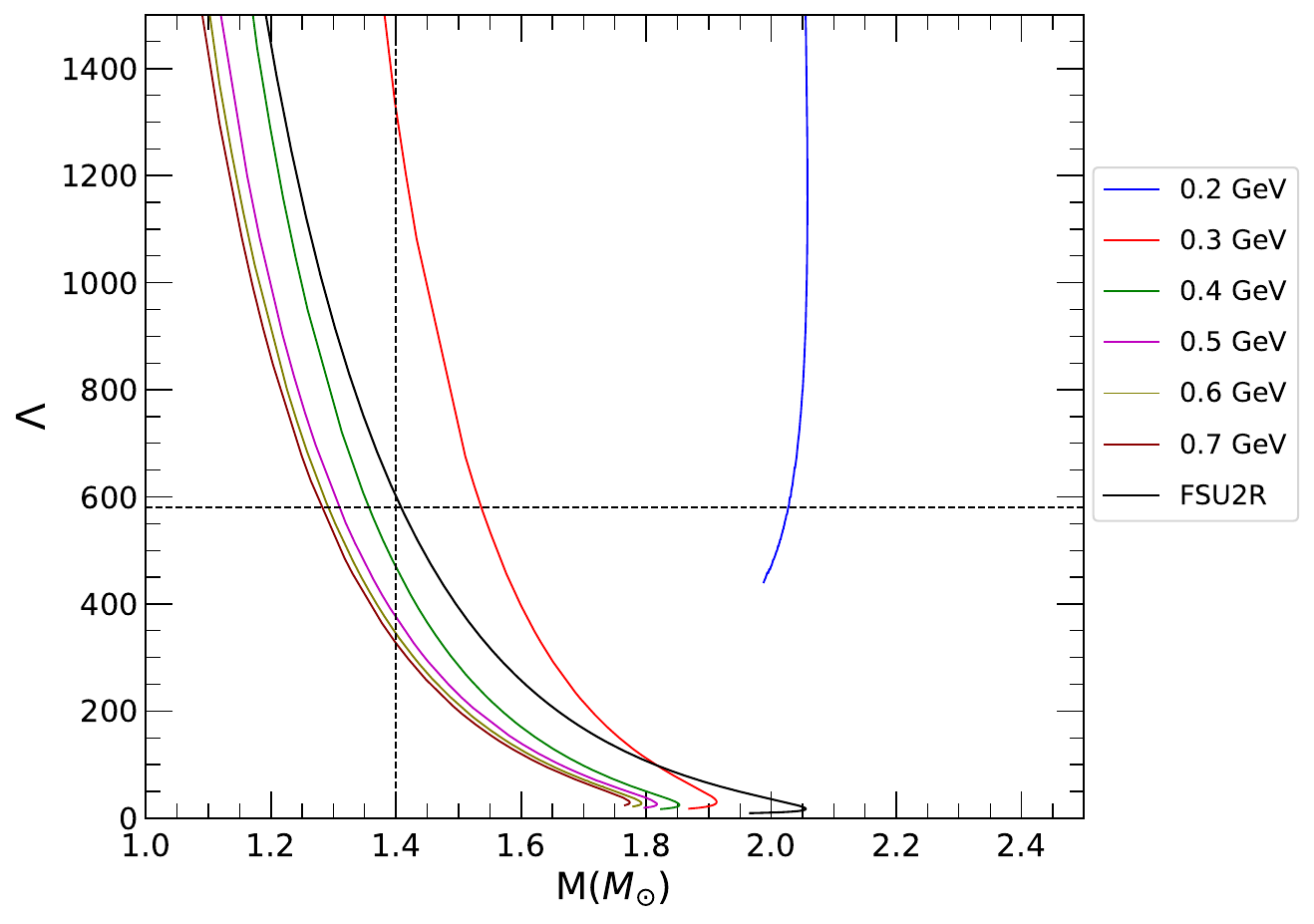}  
    \end{tabular}
    \caption{Left panel: Dimensionless tidal deformability $\Lambda$ as a function of total gravitational mass $M_T$ for various values of $\mu$, with fixed $f = 10\%$ and using the MPA1 EoS for BM. The vertical dashed line indicates $M_T = 1.4\,M_{\odot}$, while the horizontal dashed line marks $\Lambda = 580$. Right panel: Same as left panel but calculated with the FSU2R EoS for BM.}
    \label{fig:MR7}
\end{figure}

\begin{figure*}[!th]
    \centering
    \begin{tabular}{cc}
        \includegraphics[scale=0.40]{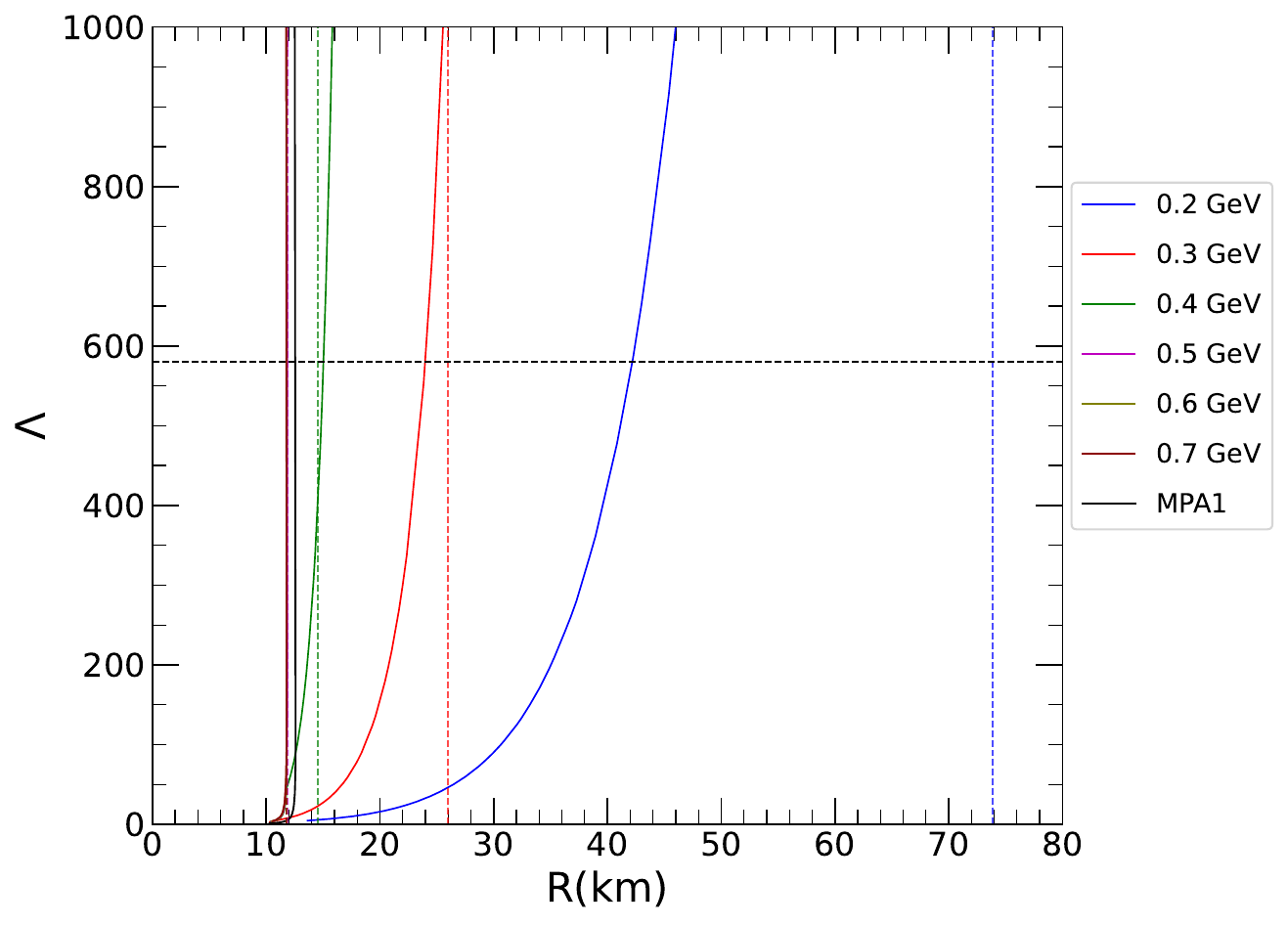} &
        \includegraphics[scale=0.40]{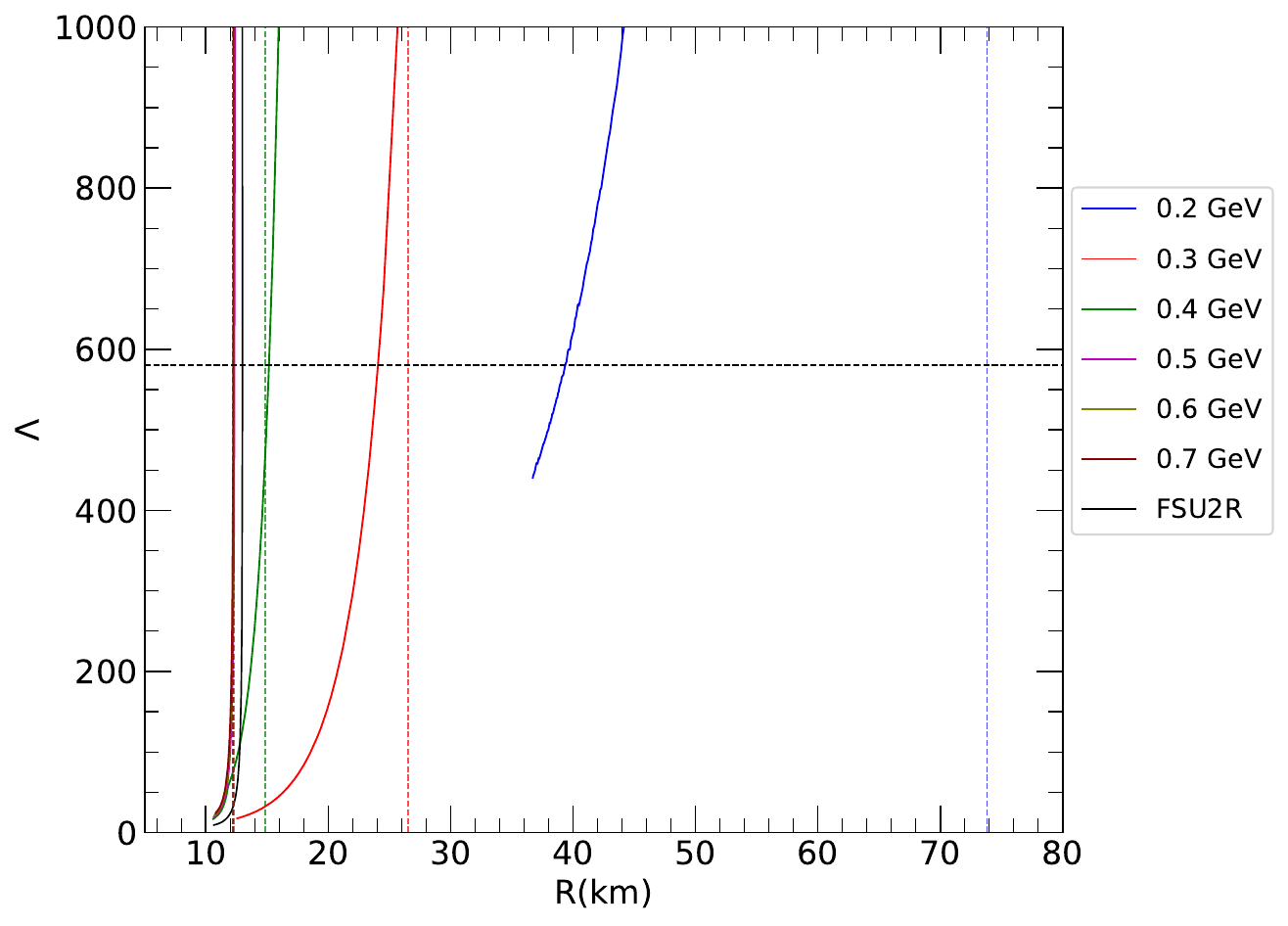}
    \end{tabular}
    \caption{Left panel: Dimensionless tidal deformability $\Lambda$ as a function of outermost radius $R$ for various values of $\mu$, with fixed $f = 10\%$ and using the MPA1 BM EoS. Every colored vertical line indicates $R_{1.4}$ derived for various values of $\mu$. Right panel: Same as left panel but calculated with the FSU2R BM EoS.}
    \label{fig:MR8}
\end{figure*}

\begin{figure*}[!th]
    \centering
    \begin{tabular}{cc}
        \includegraphics[scale=0.40]{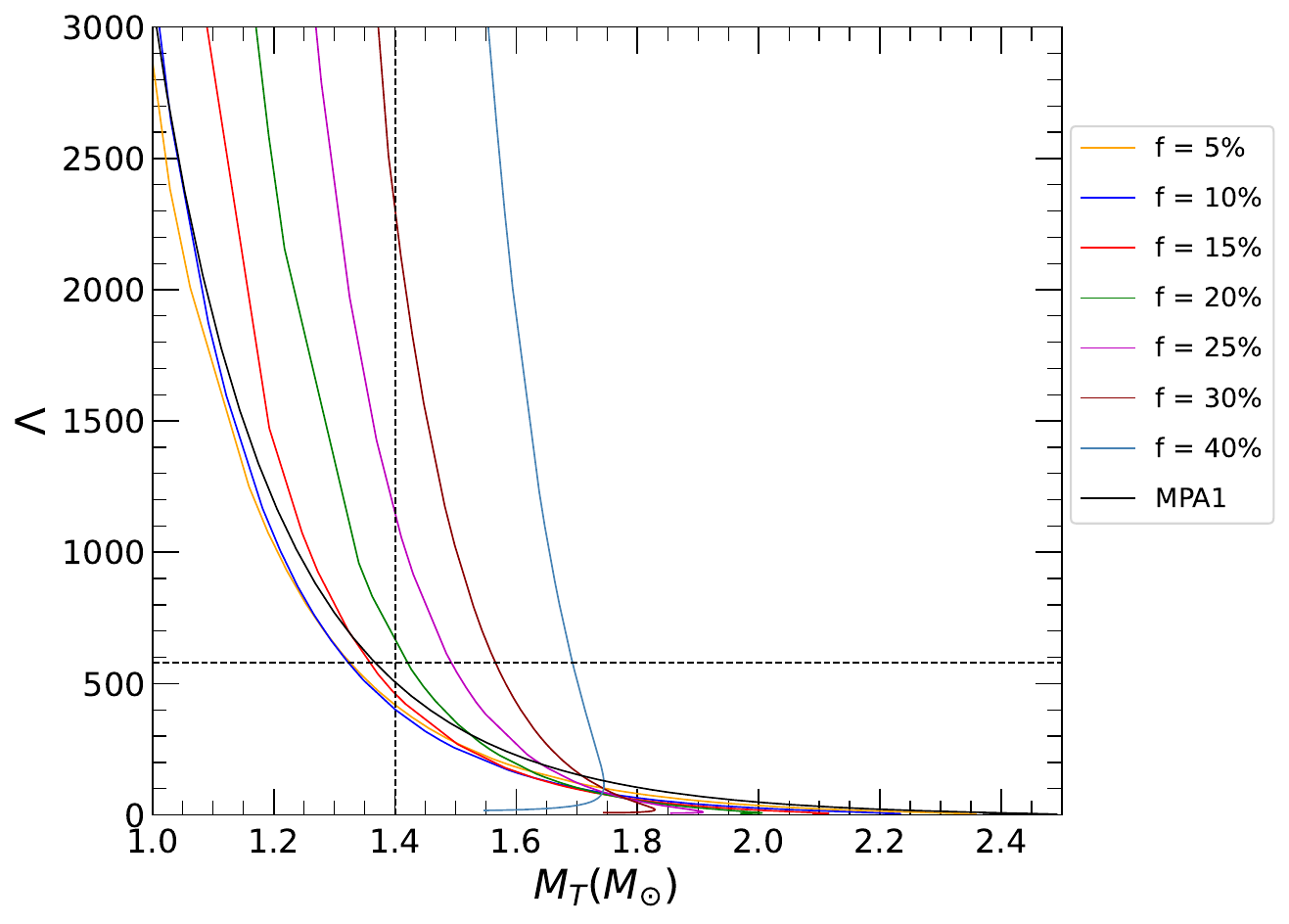} &
        \includegraphics[scale=0.40]{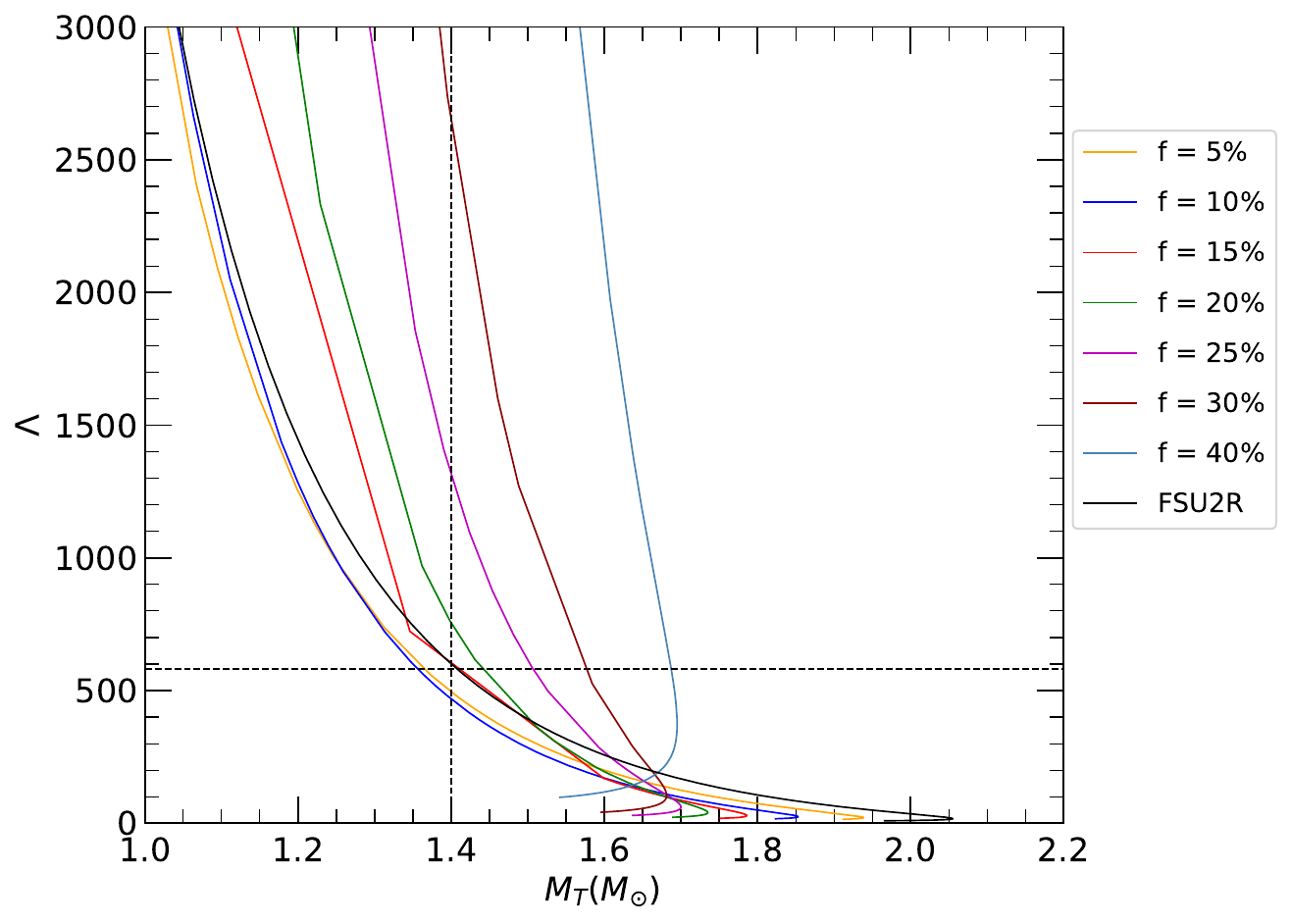}
    \end{tabular}
    \caption{Left panel: Similar to Fig.~\ref{fig:MR7} but for different values of $f$ at fixed $\mu=0.4\, GeV$ calculated with MPA1 BM EoS.Right panel: Same as left panel but calculated with the FSU2R BM EoS.}
    \label{fig:MR9}
\end{figure*}

\begin{figure*}[!th]
    \centering
    \begin{tabular}{cc}
        \includegraphics[scale=0.41]{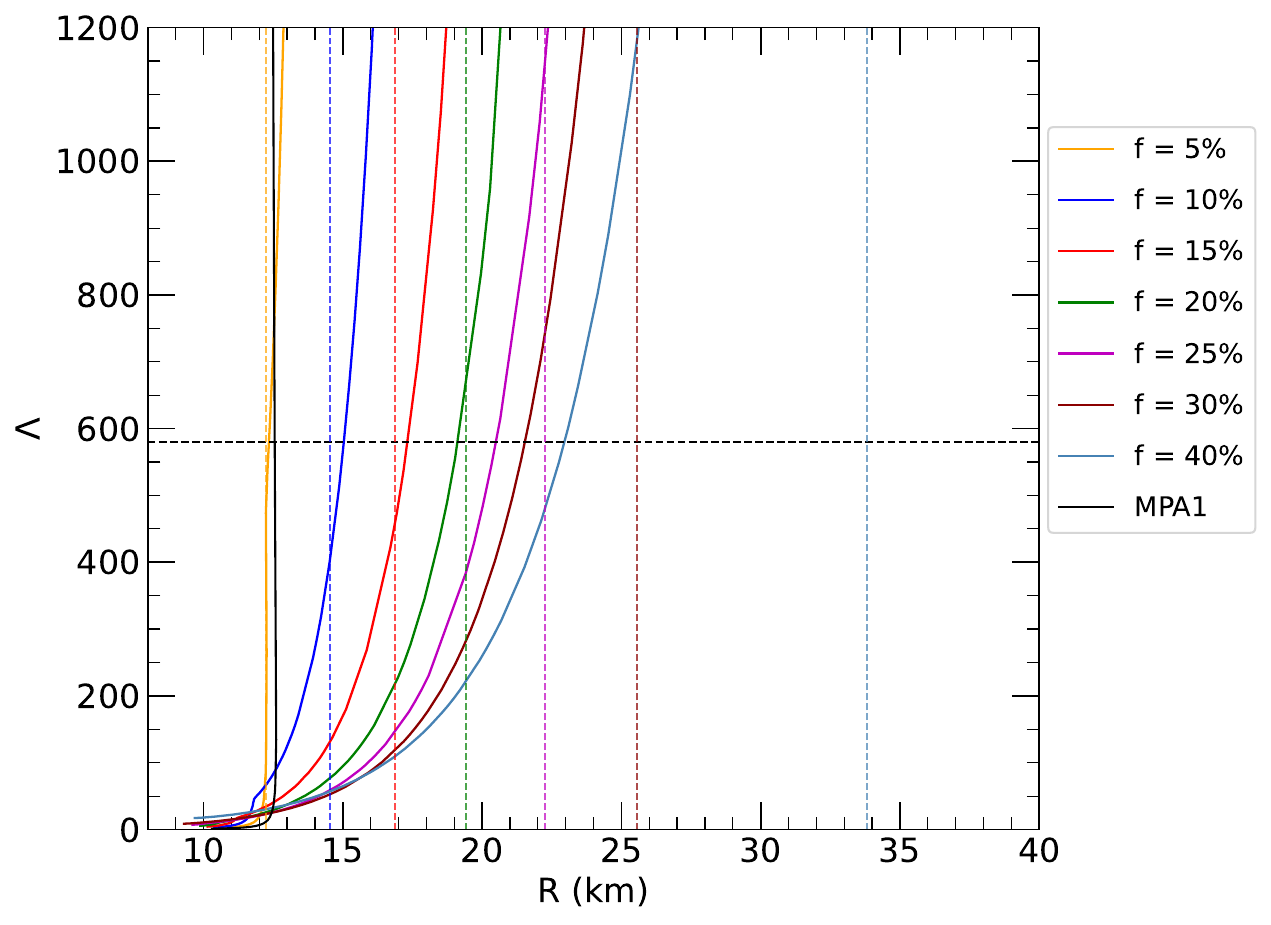} &
        \includegraphics[scale=0.41]{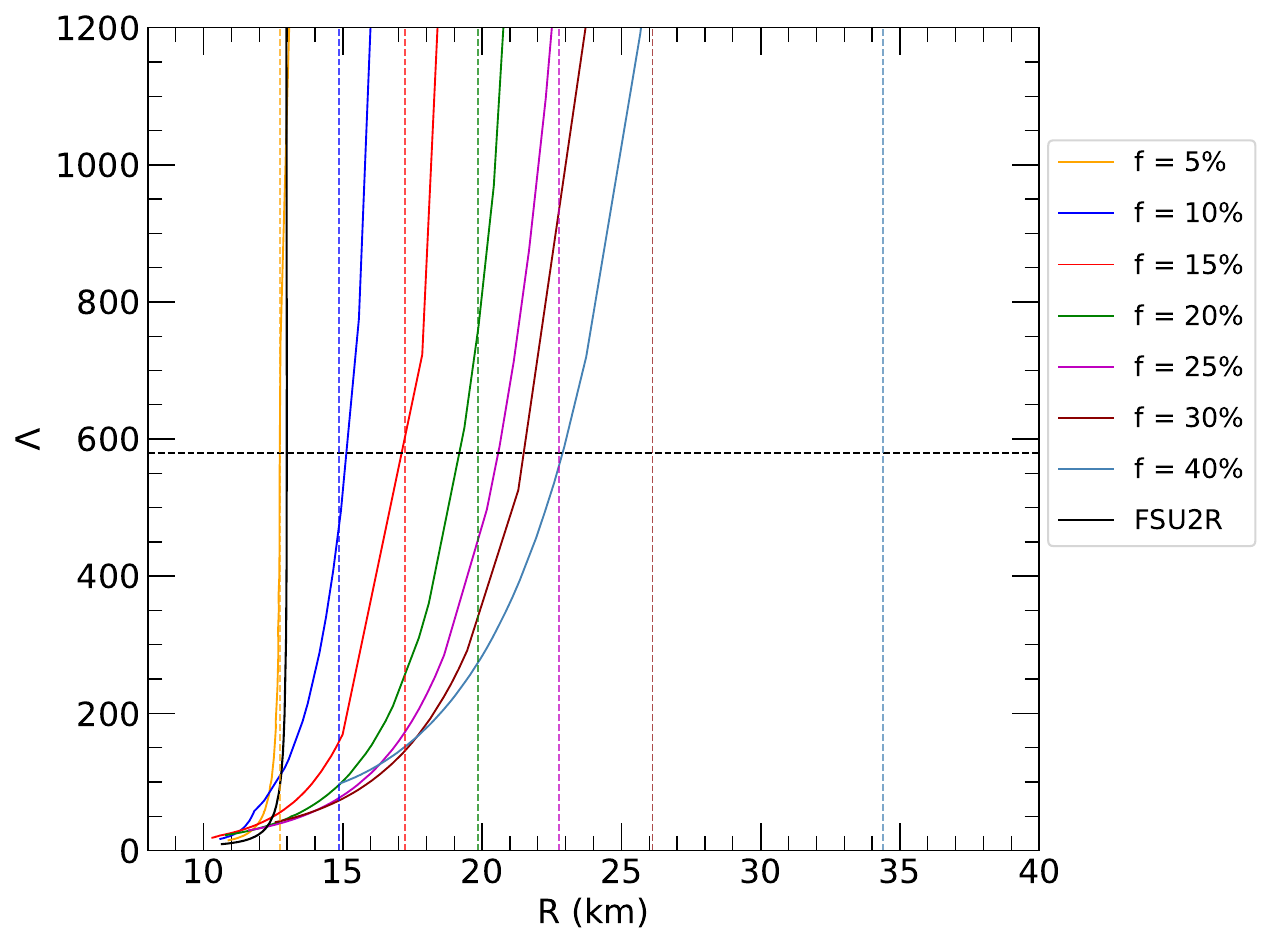}
    \end{tabular}
    \caption{Left panel: Similar to Fig.~\ref{fig:MR8} but for different values of $f$ at fixed $\mu=0.4\, GeV$ calculated with MPA1 BM EoS. Right panel: Same as left panel but calculated with the FSU2R BM EoS.}
    \label{fig:MR10}
\end{figure*}

 In this subsection we investigate how ideal Fermi gas DM—distributed either in core or halo distributions—modifies the dimensionless tidal deformability $\Lambda$. The variation of $\Lambda$ with total gravitational mass $M_T$ and outermost radius $R$ is displayed in Figs.(~\ref{fig:MR7} $-$~\ref{fig:MR10}) for different values of $\mu$ and  $f$. Horizontal black dashed lines mark the LIGO/Virgo upper limit $\Lambda_{1.4} = 580$ \citep{abbott2018gw170817}, vertical black dashed lines indicate $M_T = 1.4M_{\odot}$, and vertical colored dashed lines show radius at $1.4M_{\odot}$, $R_{1.4}$, of DANSs.$\Lambda$ for pure BM 
EoSs (MPA1 and FSU2R) is shown as solid black curves in  Figs.(~\ref{fig:MR7} $-$~\ref{fig:MR10}). For DANSs with MPA1 BM EoS, $\Lambda_{1.4}$ lies well within the LIGO/Virgo constraint (see solid black curves in left panel Fig.~\ref{fig:MR7} and left panel of Fig.~\ref{fig:MR9}),  so using the MPA1 BM EoS we can make model for both DM halo and DM core distributions without violating limits on $\Lambda$.The softer FSU2R  BM EoS produces $\Lambda_{1.4}$ slightly above the observational bound (see solid black curves in right panel Fig.~\ref{fig:MR7} and right panel of Fig. ~\ref{fig:MR9}). 

Indeed, as illustrated in Figs.(\ref{fig:MR7} $-$~\ref{fig:MR10}), $\Lambda$ generally decreases as the $M_T$ increases, while it increases with $R$. According to Eq.(~\ref{eq:ok}), $\Lambda$ depends on the ratio $\frac{R}{M_T}$; thus, its minimum value corresponds to configurations with the minimum radius and/or $M_{Tmax}$ of the DANS. The behavior of $\Lambda$ can be understood as follows: when a DM halo surrounds the NS, the $\Lambda$ increases, whereas the formation of a DM core within the star reduces $\Lambda$. This is because of  the significant dependence of 
$\Lambda$  on both the $M_T$ and  $R$ of star , as described by Eq.(~\ref{eq:ok}).

Fig.~\ref{fig:MR7} illustrates the influence of varying $\mu$ on $\Lambda$, considering a fixed value of $f\,=\,10\%$. The results show that a lower value of $\mu$ corresponds to a higher value of $\Lambda$. This trend aligns with the expected physical picture: lighter DM particles favor the formation of an extended DM halo around the star. This DM halo configuration increases $R$ of DANS, which in turn elevates the value of $\Lambda$. In the lower mass regime (\(\mu < 0.4\ \text{GeV}\)), $\Lambda$ values lie above the curve for the BM EoSs at $M_T=1.4M_{\odot}$. The formation of a DM core inside the NS (for \(\mu \geq 0.5\ \text{GeV}\)), however, causes the value of \(\Lambda\) to drop below the BM EoS predictions. A notable change in the functional behavior of $\Lambda$ curve with respect to radius \(R\) occurs at \(\mu = 0.4\ \text{GeV}\) (see both panels of Fig.~\ref{fig:MR8}). This signifies a transition from a DM halo to a DM core structure—a phenomenon analyzed extensively in the preceding section ~\ref{subsec:999}. During this transition, the star's outermost radius $R$ alternates from \(R_D\) to \(R_B\). Fig.~\ref{fig:MR8} shows that $R_{1.4}$ increases with decreasing $\mu$ as a lighter DM particle corresponds to a halo configuration. Figs.~\ref{fig:MR9} and ~\ref{fig:MR10} illustrate the influence of $f$ on
$\Lambda$

   of DANSs for a fixed value of \(\mu = 0.4\ \text{GeV}\). The results indicate that higher $f$ values facilitate the formation of a DM halo, which leads to an increase in the \(\Lambda\) values. Within this regime, the constraint \(\Lambda_{1.4} \leq 580\) is satisfied for \(f \leq 15\%\). Conversely, lower values of $f$ promote the formation of a compact DM core. This core formation subsequently induces a reduction of $\Lambda$, causing the curves to fall below the curve predicted by the BM EoSs, as demonstrated in Fig.~\ref{fig:MR9}. In Fig.~\ref{fig:MR10},  $R_{1.4}$ increases with $f$ as increasing $f$ causes formation of DM halo. 

\begin{figure*}[!th]
    \centering
    \begin{tabular}{@{}c c@{}}
        \includegraphics[width=0.567\linewidth]{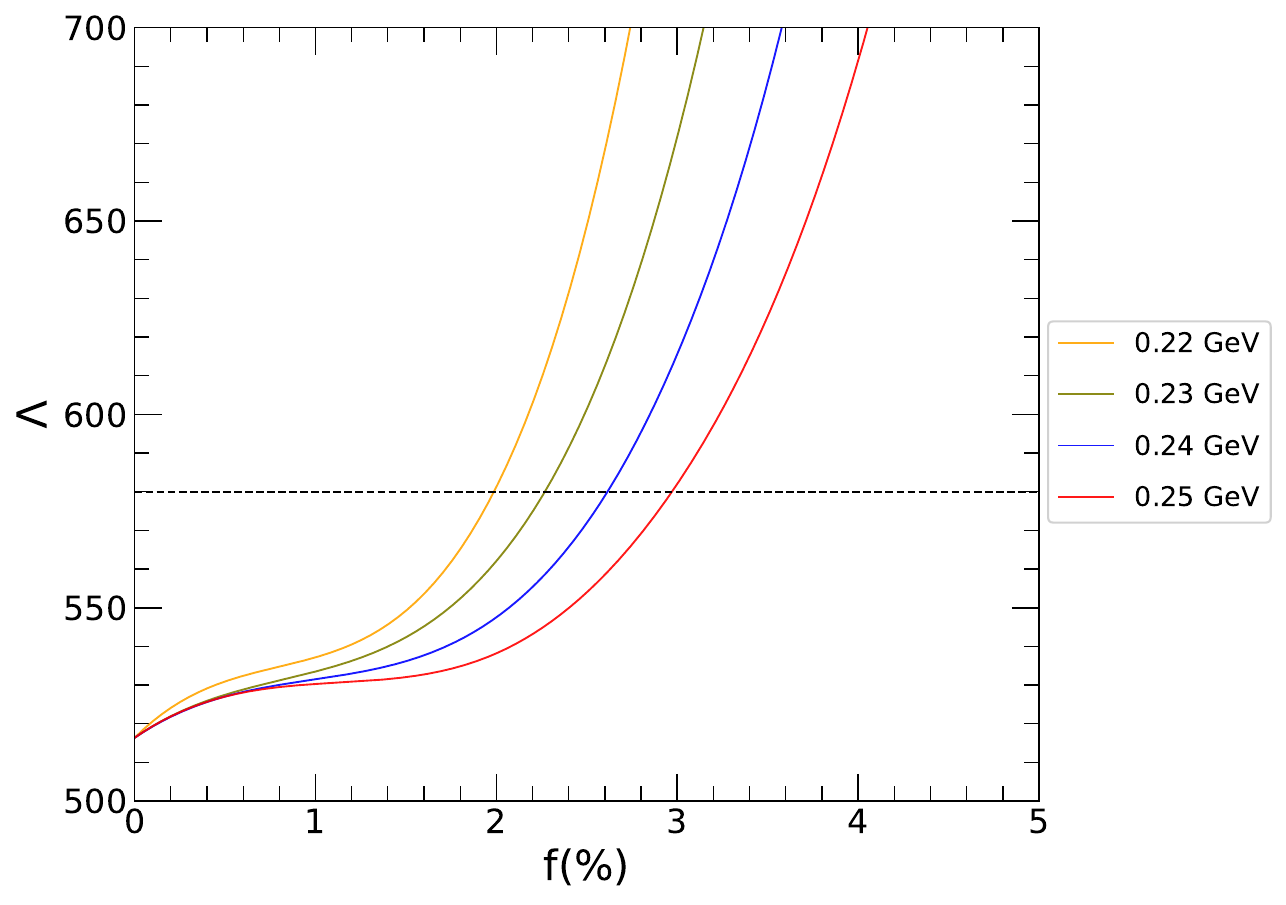} &
        \includegraphics[width=0.567\linewidth]{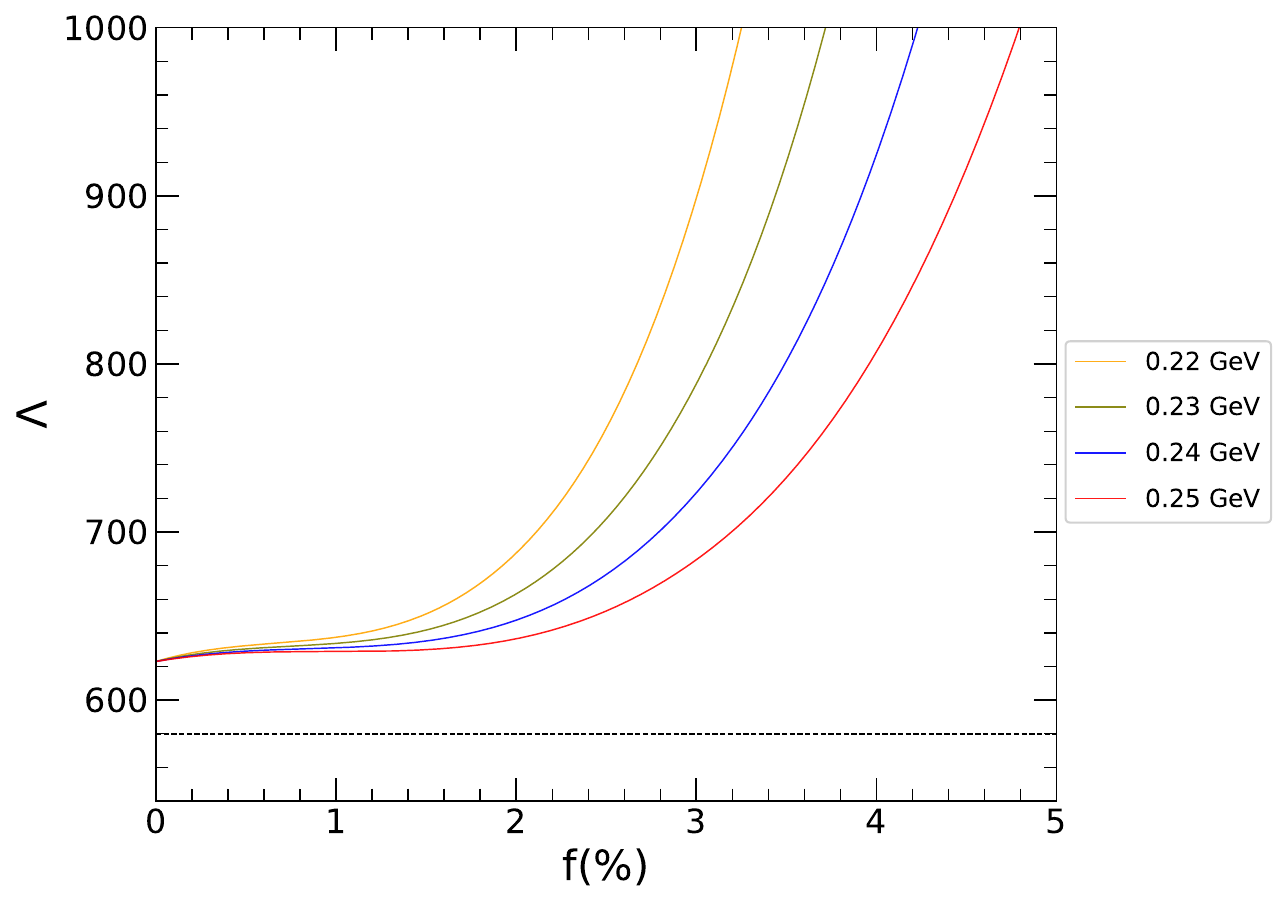} \\[12pt]
        
        \includegraphics[width=0.567\linewidth]{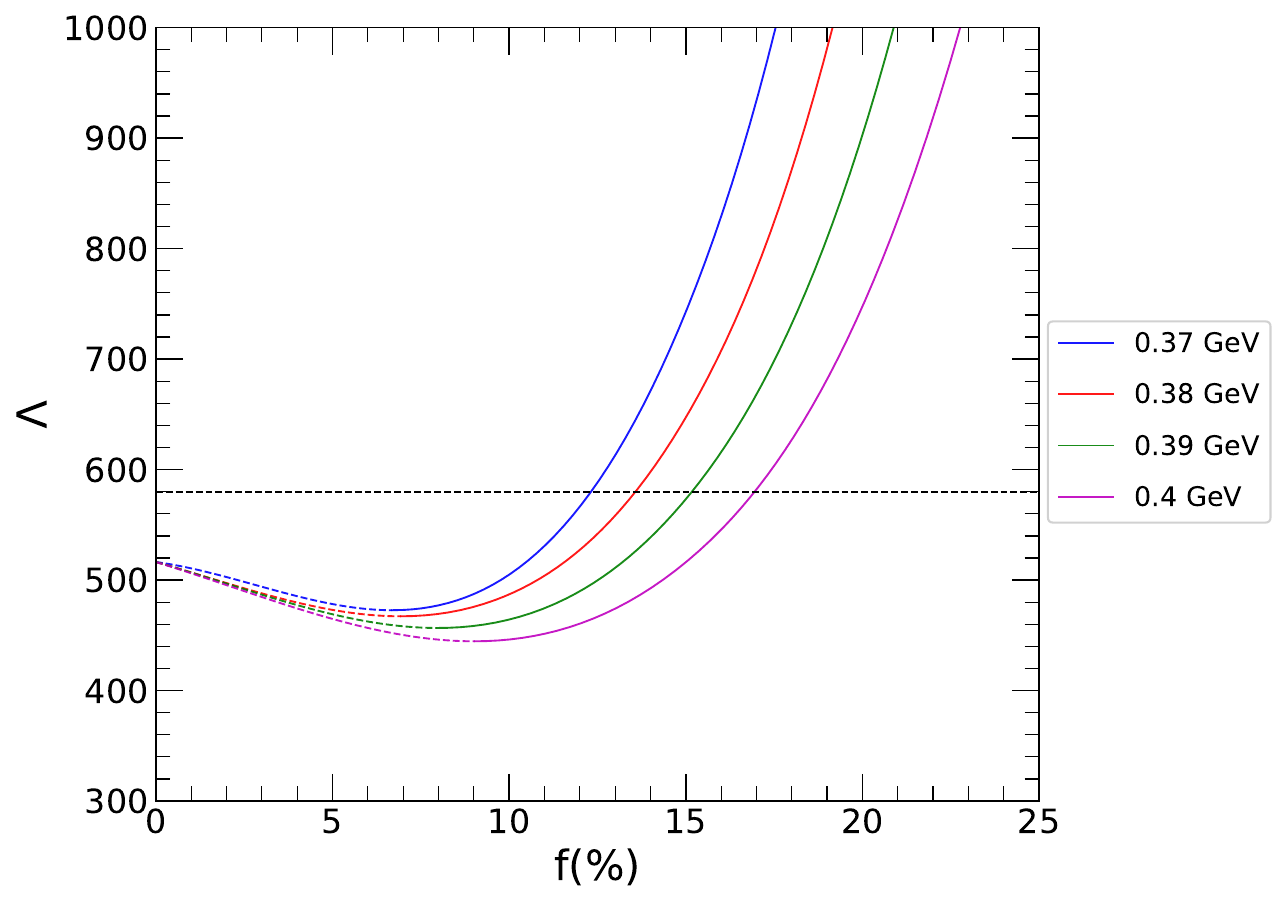} &
        \includegraphics[width=0.567\linewidth]{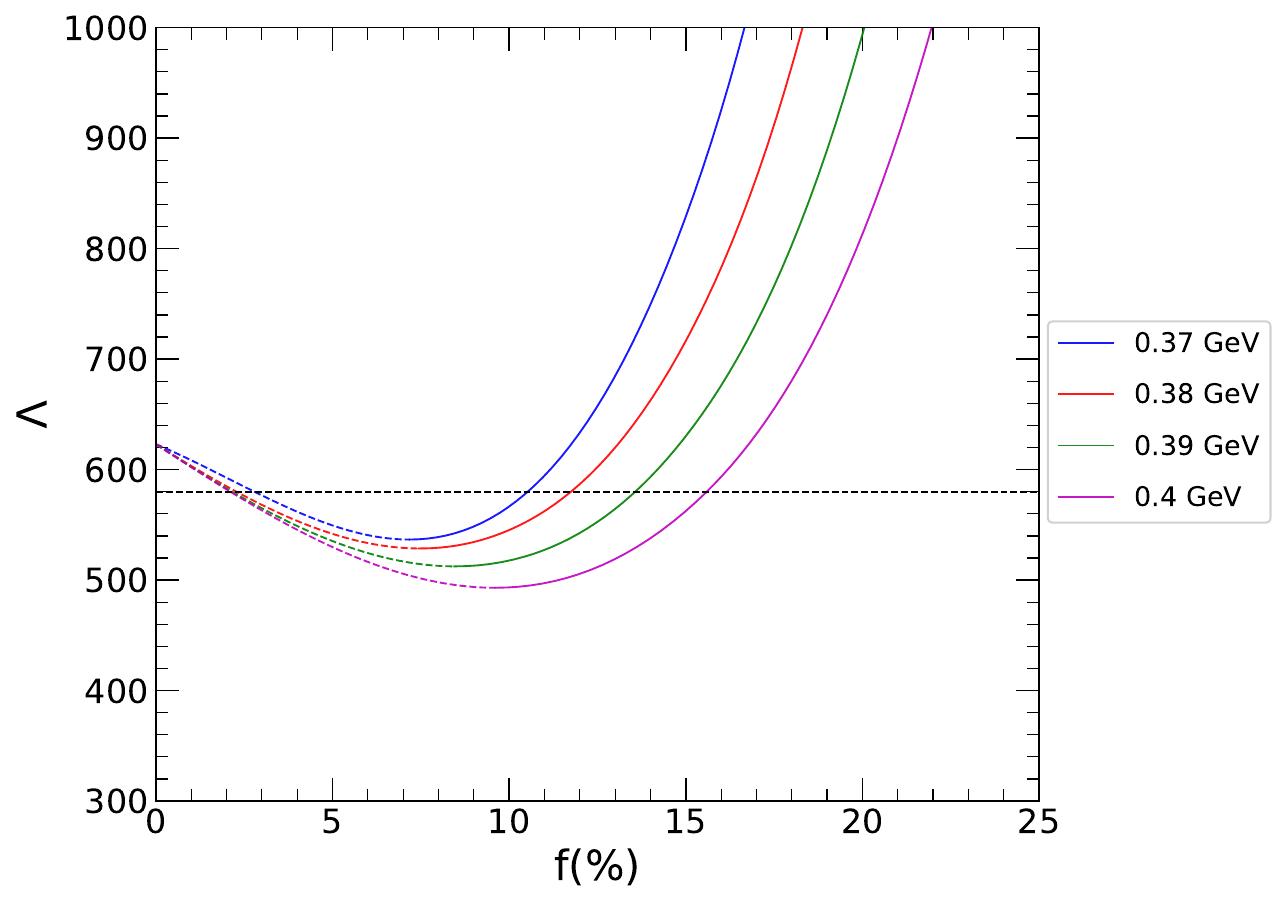}\\[12pt]
        
        \includegraphics[width=0.567\linewidth]{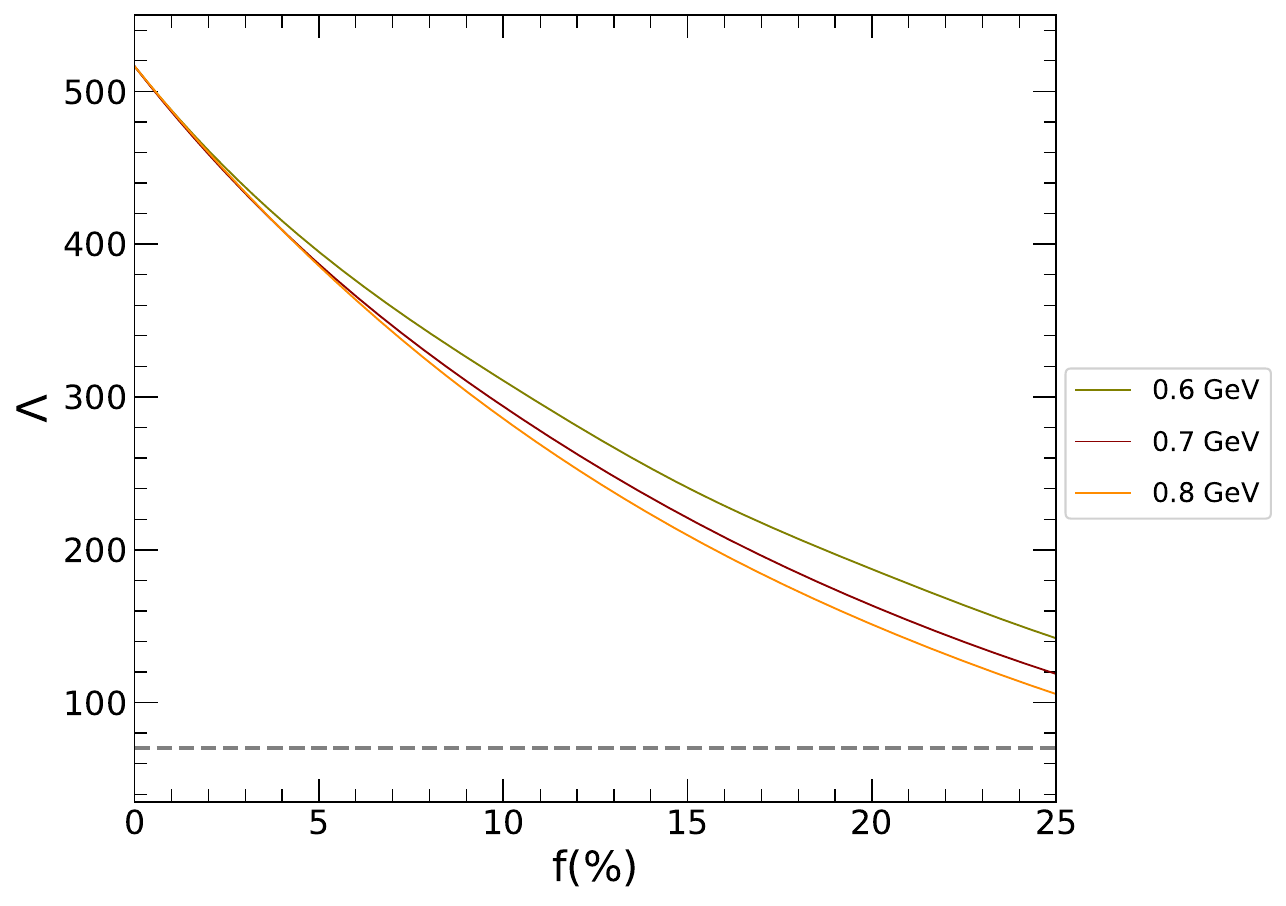} &
        \includegraphics[width=0.567\linewidth]{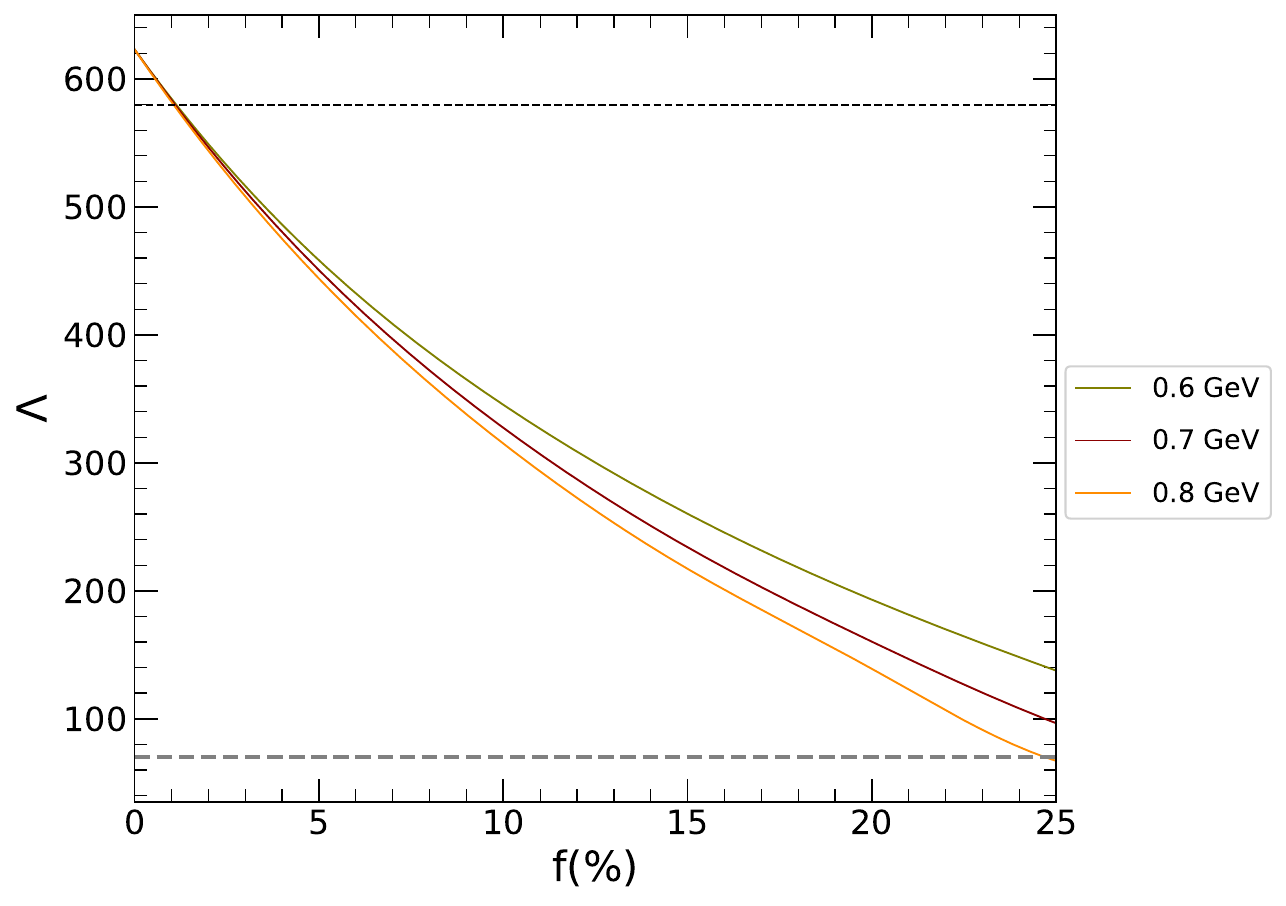}
    \end{tabular}
    \caption{Dimensionless tidal deformability $\Lambda$ is plotted as a function of  $f$ for different values of $\mu$. Upper panel: light $\mu$ values leading to DM halo formation. Middle panel: intermediate $\mu$ values showing both DM core (dashed) and halo (solid) configurations indicating transition phenomena. Lower panel: high $\mu$ values producing DM cores. All figures in left panel represent results using MPA1 BM EoS and all the figures in right panel represent results using FSU2R  BM EoS. All panels correspond to $M_T = 1.4M_{\odot}$. The black dashed line and grey dashed line indicate the $\Lambda_{1.4M_{\odot}}\,=\,580$ (LIGO/Virgo upper limit) and $\Lambda_{1.4M_{\odot}}\,=\,70$ ( LIGO/Virgo lower limit) observational constraints respectively \citep{abbott2018gw170817}.}
    \label{fig:MR11}
\end{figure*}

In Fig.~\ref{fig:MR11}, we plot $\Lambda$ as a function of $f$ for DANSs with $M_T=1.4M_{\odot}$. The upper panel shows that $\Lambda$ increases with $f$ for light DM particles, which signifies the formation a DM halo up to $f \leq 4\%$ for the MPA1 BM EoS (see upper left panel) and  up to $f \leq 4.8\%$ for FSU2R BM EoS (upper right panel). The horizontal  dashed black line represents the LIGO/Virgo upper limit of $\Lambda_{1.4} = 580$. The intersection of this line with any curve (i.e., corresponding to any $\mu$ value) determines maximum value of $f$ for which a DANS satisfies the $\Lambda_{1.4} \leq 580$ constraint. In the middle panel, we consider an intermediate $\mu$ range between 0.37 and 0.4 GeV. Here, $\Lambda$ exhibits characteristics of both DM core formation (dashed curves) and DM halo formation (solid curves). At low DM fractions—$f \leq 9\%$ for the MPA1 EoS (middle left panel) and $f \leq 10\%$ for the FSU2R EoS (middle right panel)—DM condenses into the NS core, causing $\Lambda$ to decrease with increasing $f$. The minimum $\Lambda$ occurs at a specific $f$ value, indicating the core–halo transition point, after which $\Lambda$ rises with $f$ in the DM halo regime. In the lower panel of Fig.~\ref{fig:MR11}, we present results for massive fermions, which generate dense DM cores. Here, $\Lambda$ decreases monotonically with $f$, with the reduction being more pronounced for larger $\mu$ at a fixed $f$ value \citep{abbott2018gw170817}. Here we must care about lower observational limit of  LIGO/Virgo $\Lambda_{1.4}\,=\,70$ \citep{abbott2018gw170817} (shown by grey dashed lines in lower panel). The intersection of this line with any curve (i.e., corresponding to any $\mu$ value) also specifies the maximum value of $f$ for which a DANS satisfies the constraint $\Lambda_{1.4}\geq 70$.
\section{Combined EM and GW Observational Constraints}
\label{77777}
\begin{figure}[!th]
    \centering
    \begin{tabular}{cc}
       \includegraphics[scale=0.44]{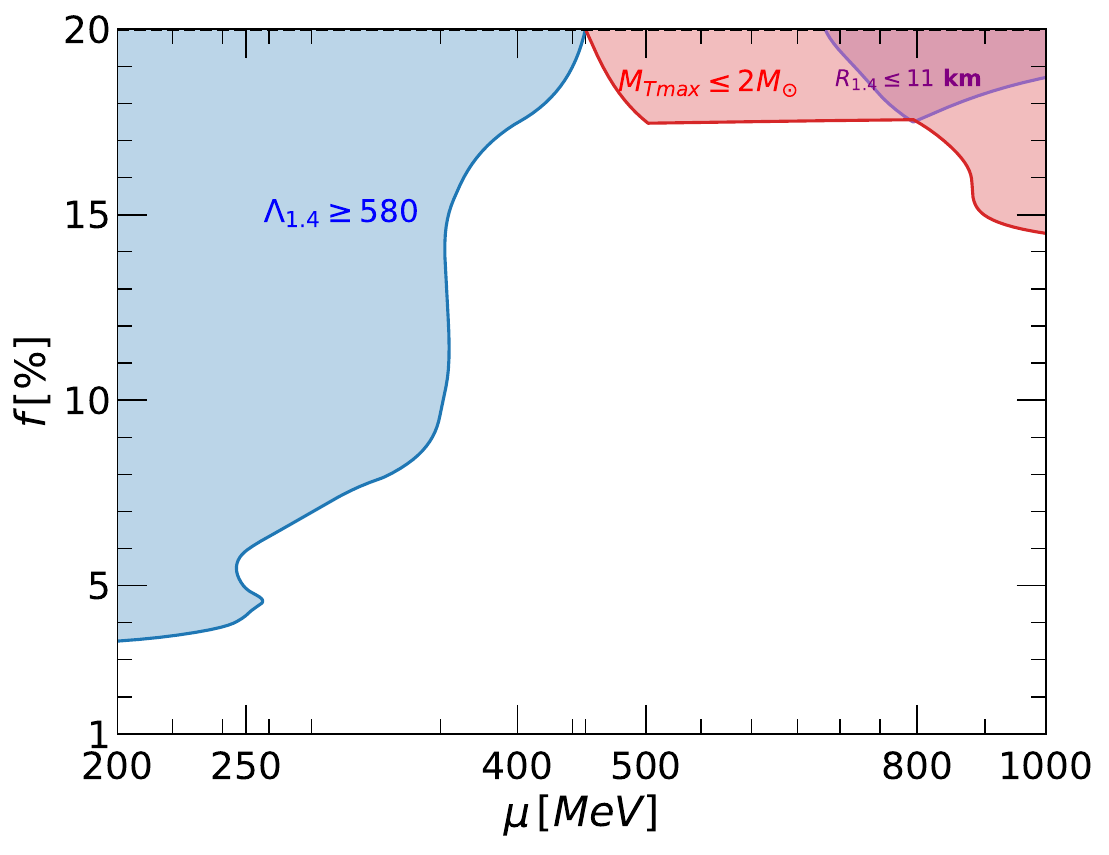}  &
        \includegraphics[scale=0.44]{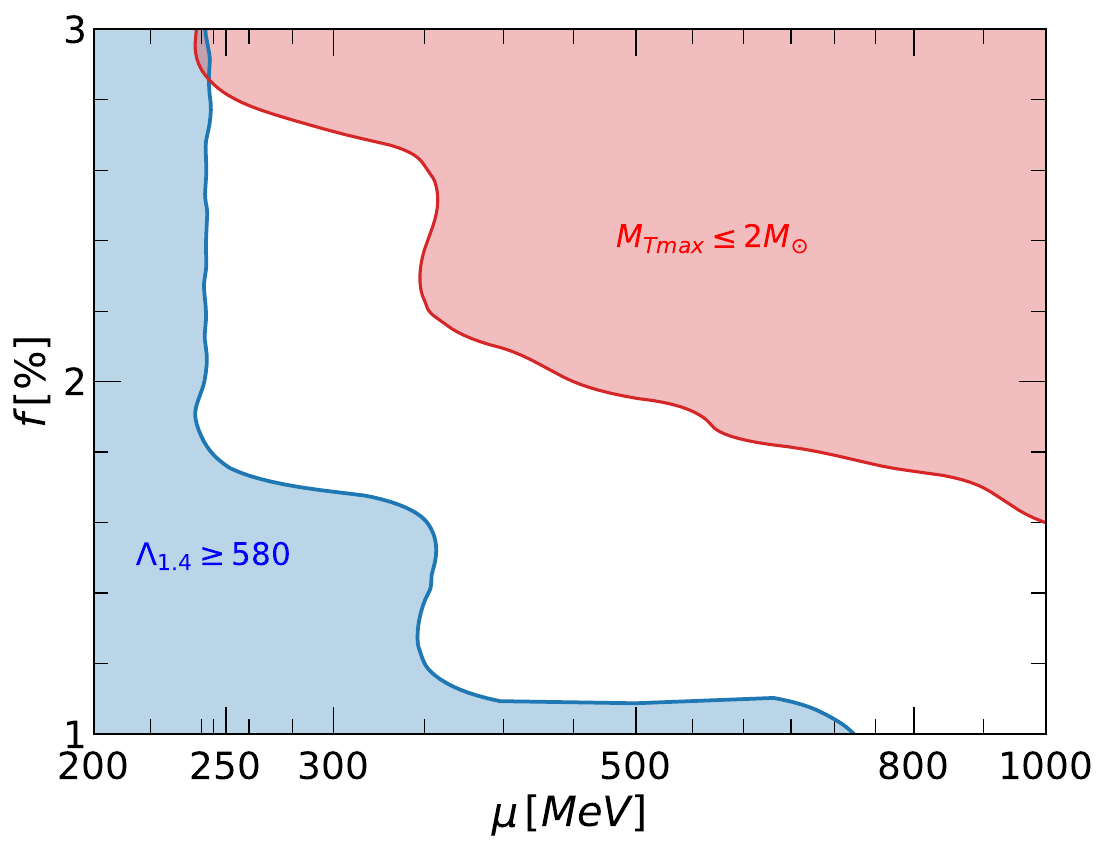}  
    \end{tabular}
    \caption{The excluded regions in the $f-\mu$ parameter space are shown with distinct colors for the MPA1 (left panel) and FSU2R (right panel) BM EoSs. The red area represents configurations where  $M_{Tmax}\,\leq\,2M_{\odot}$. The blue region corresponds to $\Lambda_{1.4} \geq 580$ for tidal deformability of DANSs  with $M_T=1.4M_{\odot}$. The purple zone corresponds to DANS  with  observable radius $R_{1.4} \leq 11$ km.}
    \label{fig:MR12}
\end{figure}

In this section, we investigate the parameter space of the fermionic DM model by considering three multimessenger observables: the visible (BM) radius ($R_B$), the maximum total gravitational mass ($M_{Tmax}$), and the tidal deformability ($\Lambda$) of DANSs. 
Two EoSs, MPA1 and FSU2R, are employed to model the baryonic component of the DANSs. We perform a systematic parameter scan over the DM model parameters within the ranges $\mu \in [200, 1000]~\mathrm{MeV}$ and $f \in [1,20]\%$ for the MPA1 EoS and $f \in [1,3]\%$ for the FSU2R EoS, while imposing astrophysical constraints from NS observations: $M_{\mathrm{Tmax}} \geq 2M_{\odot}$, $R_{1.4} \geq 11~\mathrm{km}$, and $\Lambda_{1.4} \leq 580$.The presence of DM within or around a NS can significantly alter its astrophysical properties. Specifically, the formation of a DM core tends to substantially reduce the star's $M_{Tmax}$, $R_B$ and $\Lambda$ potentially leading to violations of established observational bounds. On the other hand, a DM halo configuration surrounding the star may result in a moderate increase in the $M_{Tmax}$ and $\Lambda$---a scenario that may conflict with existing tidal deformability limits. Fig.~\ref{fig:MR12} presents the results of a systematic scan over the $f$-$\mu$ parameter space for the MPA1 and FSU2R BM EoSs, incorporating astrophysical constraints derived from NICER and LIGO/Virgo observations. The white regions correspond to parameter combinations that satisfy all imposed constraints: $M_{\mathrm{Tmax}} \geq 2M_{\odot}$, $R_{1.4} \geq 11~\mathrm{km}$, and $\Lambda_{1.4} \leq 580$. As shown in the left panel (for DANSs with MPA1 BM EoS), the constraint on the $M_{Tmax}$ provides a more stringent limitation than the $R_{1.4}$ constraint. For low DM particle masses ($\mu \lesssim 350~\mathrm{MeV}$), the constraint on the tidal deformability $\Lambda_{1.4}$ rules out the existence of a stable, substantial DM component within a NS. However, for more massive fermions, the allowed value of $f$ increases significantly, with its upper limit eventually determined by the combined effect of these three astrophysical constraints. Satisfying all three constraints simultaneously yields a maximum permissible DM fraction $f$ of approximately $20\%$ at  $\mu \approx 440~\mathrm{MeV}$.

In the right panel of Fig.~\ref{fig:MR12}, the FSU2R EoS is employed for the BM component. Our analysis shows that the parameter region corresponding to $\mu\,\leq240\,\text{MeV}$ is ruled out by the $\Lambda_{1.4}$ constraint for all considered values of $f$. For $\mu$ values between $240$ and $350\,\text{MeV}$, the region corresponding to $f\lesssim 1.8\%$ is excluded by $\Lambda_{1.4}$ constraint. Similarly, for $\mu$ in  the range $350$–$700\,\text{MeV}$, the region corresponding to $f\lesssim 1.2\%$ is disfavored by LIGO/Virgo limit on $\Lambda_{1.4}$. For $\mu>700$ MeV the region for $f\geq\,1\%$ is consistent with this $\Lambda_{1.4}$ constraint. This pattern arises because the tidal deformability of the pure FSU2R EoS, $\Lambda_{1.4} \approx 622$, exceeds the current observational upper limit of $\Lambda_{1.4} \lesssim 580$. The inclusion of a DM core reduces $\Lambda$ for massive fermions and above a certain value of $f$, the resulting $\Lambda_{1.4}$ becomes consistent with the observational bound. Conversely, for lighter fermions, DM halo formation increases $\Lambda$, which is inconsistent with GW constraints. Notably, the radius constraint $R_{1.4}$ does not impose any additional limitations on the $f$–$\mu$ parameter space and the $M_{Tmax}$ constraint predominantly excludes the upper-right region of the parameter space. A maximum allowed DM fraction of $f \approx 2.85\%$ is found near $\mu \approx 240\,\text{MeV}$, consistent with all three observational bounds. Thus, compatibility with the aforementioned constraints is achieved for  $\mu \gtrsim 240\,\text{MeV}$ and for $f$ in the range $[f_{\mathrm{th}}, 2.85]\%$, where $f_{\mathrm{th}}$ represents the threshold DM fraction above which models are consistent with all three astrophysical constraints considered here. The value of $f_{\mathrm{th}}$ varies for different ranges of $\mu$ as seen in the right panel of Fig.~\ref{fig:MR12}.

In summary, we find that DANS models with the stiff MPA1 BM EoS—which only marginally satisfies observational limits—place tighter constraints on the fermionic DM parameter space than  models with the softer FSU2R BM EoS. While the formation of a DM core or halo, and its consequent influence on observable NS properties, is independent of the BM EoS, the specific constraints on the DM model are sensitive to the underlying nuclear model. This sensitivity arises because different baryonic EoSs predict distinct values for bulk stellar properties. For both EoSs considered, lighter DM particles are favored by $M_{Tmax}$ and $R_{1.4}$ constraints across the entire range of $f$. However, in this low-mass regime, the $\Lambda_{1.4}$ bound significantly restricts the permissible value of $f$. The maximum allowed value of $f$ is approximately $20\%$ for the model with the MPA1 BM EoS. On the other hand, for the model with FSU2R BM EoS, the allowed value of $f$ is confined to a range between a minimum threshold, $f_{\mathrm{th}}$, and $2.85\%$.
\section{Conclusion and Remarks}
\label{88888}
In this work, we have modeled DANSs by considering a zero-temperature ideal Fermi gas for the DM component, alongside the BM EoSs MPA1 and FSU2R. Our analysis demonstrates that the spatial distribution of DM—whether forming a central core or an extended halo within the mixed compact object—depends critically on the DM model parameters, namely the fermion mass $\mu$ and the fractional amount of DM $f$. While the two selected BM EoSs encompass a range of nuclear matter properties inferred from observations, we find that the general phenomenology of DM core versus halo formation is not sensitive to the specific BM model.By examining the equilibrium configurations of DANSs, we have probed the DM parameter space in light of recent astrophysical constraints,including the visible (BM) radius ($R_B$), the total maximum mass ($M_{Tmax}$), and the  dimensionless tidal deformability ($\Lambda$). We show that when DM condenses into a core, it reduces $M_{Tmax}$, $R_B$, and $\Lambda$ compared to a pure baryonic NS. Conversely, the presence of DM in a halo surrounding the NS increases these observable quantities. Given that core formation diminishes both $M_{Tmax}$ and $R_B$, our results indicate that lighter DM fermions and lower DM fractions are favored by recent NICER measurements of PSR J0030+0451 and PSR J0740+6620.

In light of the joint NICER and LIGO/Virgo constraint indicating a lower limit of approximately $11$ km for the radius of a $1.4M_{\odot}$ NS, we have analyzed the variation of  $R_B$ and the DM radius $R_D$ with respect to the parameters $f$ and $\mu$. Our results show that as $f$ increases, $R_B$ is primarily a decreasing function for both core and halo configurations, while $R_D$ exhibits a consistent increase. For all values of $\mu$ considered here, a transition from DM core configuration ($R_B>R_D$) to DM halo configuration ($R_D>R_B$) would occur at a specific value of $f$, where the outermost radius ($R$) of the DANSs changes from $R_B$ to $R_D$. Furthermore, we find that for massive fermions, $R_B$ is significantly reduced and can fall below the $11$ km, a feature that can be used to place further constraints on the DM model parameter space. Future X-ray observatories, including STROBE-X \citep{ray2019strobe}, ATHENA \citep{barcons2012athena}, and eXTP \citep{watts2019dense}, will possess the sensitivity required to detect the predicted variation rate in $R_B$ of a DANS.

Finally, we perform a comprehensive scan over $\mu$ and $f$ across specified ranges. In this analysis, we incorporate the combined multi-messenger astrophysical constraints for NSs derived from GW and EM observations: $M_{\text{Tmax}} \geq 2M_{\odot}$, $R_{1.4} \geq 11\ \text{km}$ and $\Lambda_{1.4} \leq 580$. We observe that $M_{Tmax}$ and $R_{1.4}$ limits primarily constrain scenarios involving more massive  fermions, with the $M_{\text{Tmax}}$ condition providing the broader exclusion. In contrast, $\Lambda_{1.4}$ bound predominantly restricts lighter fermions, a consequence of DM halo formation. Our results reveal a maximum allowed value of $f$ of approximately $2.85\%$ for DANSs with the FSU2R  BM EoS and approximately $20\%$ for DANSs with MPA1 BM EoS, indicating that any NS consistent with these observational limits can contain only a relatively small amount of DM. \textcolor{blue}{ In this context, recent Bayesian studies on fermionic DANS have similarly demonstrated that current astrophysical observations primarily constrain the DM fraction, while providing limited information on the particle mass or coupling \citep{arvikar2025exploring}} Furthermore, for the FSU2R baryonic model, we demonstrate that for fermion masses $\mu \geq 240\ \text{MeV}$, the allowed parameter region corresponds to $f \gtrsim f_{\text{th}}\%$, where $f_{\text{th}}$ represents a minimum threshold fraction.

This study can be extended by incorporating multiple BM and DM EoSs. Bayesian analysis, using the latest multi-messenger observational data, could then be used to identify the optimal parameter space. In the present work, we have examined the general influence of an ideal Fermi gas DM component on NS properties. We explored a wide DM parameter space using two substantially different BM EoSs. However, uncertainties in the BM EoS at high densities, combined with the fact that DM presence could mimic certain BM behaviors, prevent us from drawing definitive conclusions. Nevertheless, applying multiple independent observational constraints to these exotic composite objects may help break the degeneracies in modeling the internal structure of compact stars. 

DANSs can provide alternative explanations for several exotic observations reported in recent years. For example, the secondary component of the GW190814 event, with a mass of $\sim 2.6\,M_{\odot}$ \citep{abbott2020gw190814,di2022can,lee2021could}, could be explained by this framework. Additionally, NICER collaboration measurements present an interesting puzzle. PSR J0740+6620 has a gravitational mass of $2.08 \pm 0.07~M_{\odot}$ and a radius of $12.35 \pm 0.75$ km \citep{miller2021radius}. This radius value is comparable to inferred radius of the less massive pulsar PSR J0030+0451 ($M \sim 1.4~M_{\odot}$), reported as $12.45 \pm 0.65$ km \citep{miller2021radius}. This weak dependence of radius on mass challenges standard NS theories and mass-radius relationships \citep{christian2022confirming,lin2024indication,drischler2022large,legred2021impact}. The presence of a DM component could potentially resolve this tension. More recently, the HESS collaboration observed an exceptionally light and compact object in the supernova remnant HESS J1731-347. It has a mass of $0.77^{+0.20}_{-0.17}~M_{\odot}$ and a radius of $10.4^{+0.86}_{-0.78}$ km \citep{doroshenko2022strangely}. Such a light compact object is also consistent with DANS models \citep{routaray2023dark,sagun2023nature}.

Ongoing observations from the LIGO/Virgo/KAGRA collaboration \citep{abbott2021observation}, LISA \citep{baker2019laser}, the Einstein Telescope \citep{maggiore2020science}, and Cosmic Explorer \citep{evans2021horizon} will provide numerous binary merger detections. This data will be crucial for investigating the properties of DANSs \citep{baryakhtar2022dark,evans2023cosmic}. To accurately characterize these objects, the presence of a DM core or halo must be properly incorporated into numerical relativity simulations. This is essential for reliably computing the GW spectrum and waveforms emitted during the merger and post-merger phases of binary systems containing at least one DANS \citep{ellis2018search,giudice2017hunting,emma2022numerical,bauswein2023compact}. Other observational techniques can also probe the DM content in and around stars. For example, a DM halo surrounding a NS could cause gravitational microlensing, leading to measurable changes in the brightness of a background source. This signature could help distinguish DANSs from other dense objects. Similarly, an extended halo may produce a detectable effect on the Shapiro time delay. Our analysis, based on an ideal Fermi gas model for DM, provides a foundational framework for this exploration. Ultimately, precise measurements of compact object properties from upcoming astrophysical instruments may reveal the nature of DM and test the hypothesis of its existence within NSs.

\section*{Acknowledgements}
M. Kalam thanks ICARD, Aliah University for providing research facilities and the Inter-University Centre for Astronomy and Astrophysics (IUCAA), Pune, India, for providing the Associateship programme. M. Kalam is grateful to RDC, Aliah University for funding (Project ID: AURDC/2024-25/PHY/07), in part, for this research project. M. Murshid and M.Molla thank Inter-University Centre for Astronomy and Astrophysics (IUCAA), Pune, India, for providing the research facilities during the visit to IUCAA. M. Molla also thanks the University Grants Commission (UGC) for providing financial support through the Senior Research Fellowship (SRF).

\section*{Data Availability Statement}
This manuscript has no associated data or the data will not be deposited. [Authors' comment: This manuscript has no measured data associated; the plots involve data generated by modelling.]

\section*{Conflict of interest}
The authors declare no conflict of interest.

\bibliographystyle{elsarticle-num}
\bibliography{references}
\end{document}